\documentclass[%
 reprint,
 amsmath,amssymb,
 aps,
]{revtex4-2}

\usepackage{graphicx}
\usepackage{dcolumn}
\usepackage{bm}

\usepackage[colorlinks=true,linkcolor=blue,anchorcolor=black,citecolor=blue,filecolor=blue,menucolor=blue,runcolor=blue,urlcolor=blue]{hyperref}

\usepackage{xcolor, soul}
\sethlcolor{green}
\usepackage{braket}
\usepackage{subfigure}
\usepackage{comment}
\usepackage{amsmath}

\begin{document}

\preprint{APS/123-QED}

\title{Long-distance photon-mediated and short-distance entangling gates in three-qubit quantum dot spin systems}
\author{Nooshin M. Estakhri}%
 \email{estakhri@umich.edu}
\author{Ada Warren}
\author{Sophia E. Economou}%
 \email{economou@vt.edu}
 \author{Edwin Barnes}%
 \email{efbarnes@vt.edu}
\affiliation{%
 Department of Physics, Virginia Polytechnic Institute and State University, 24061 Blacksburg, VA, USA
}%
\affiliation{
 Virginia Tech Center for Quantum Information Science and Engineering, Blacksburg, VA 24061, USA
}%

\date{\today}

\begin{abstract}
Superconducting resonator couplers will likely become an essential component in modular semiconductor quantum dot (QD) spin qubit processors, as they help alleviate cross-talk and wiring issues as the number of qubits increases. Here, we focus on a three-qubit system composed of two modules: a two-electron triple QD resonator-coupled to a single-electron double QD. Using a combination of analytical techniques and numerical results, we derive an effective Hamiltonian that describes the three-qubit logical subspace and show that it accurately captures the dynamics of the system. We examine the performance of short-range and long-range entangling gates, revealing the effect of a spectator qubit in reducing the gate fidelities in both cases. We further study the competition between non-adiabatic errors and spectator-associated errors in short-range operations and quantify their relative importance across practical parameter ranges for short and long gate times. We also analyze the impact of charge noise together with residual coupling to the spectator qubit on inter-module entangling gates and find that for current experimental settings, leakage errors are the main source of infidelities in these operations. Our results help pave the way toward identifying optimal modular QD architectures for quantum information processing on semiconductor chips.
\end{abstract}

\maketitle


\section{\label{sec-l1:Introduction}Introduction}

Encoding qubits in the spin of electrons in electrostatically defined quantum dots (QDs), following the Loss-DiVincenzo proposal~\cite{loss1998quantum}, continues to be one of the most promising platforms for future large-scale quantum processors~\cite{de2021materials, burkard2021semiconductor}. Among the different types of QD-based quantum processors, silicon (Si)-based systems are especially promising due to their prospects for scalability and compatibility with existing semiconductor manufacturing processes~\cite{veldhorst2017silicon, ha2021flexible,Zwerver_NatElectron_2022}. Additionally, initialization, logical operations, and measurements can be conducted with all-electrical control signals~\cite{blumoff2022fast, weinstein2023universal}. Very high single-qubit gate fidelities $\left(>99.9\%\right)$ have been experimentally demonstrated for these qubits~\cite{xue2022quantum, takeda2021quantum, yang2019silicon, yoneda2018quantum, mills2022two, tanttu2023stability}. Also, two-qubit gates with fidelities higher than the $99\%$ threshold for certain quantum error correcting codes~\cite{fowler2012surface} have been experimentally realized across different silicon-based QD platforms~\cite{xue2022quantum, noiri2022fast, mills2022two,tanttu2023stability}, with state preparation and measurement (SPAM) fidelities exceeding $97\%$~\cite{mills2022two, blumoff2022fast}. Additionally, these processors can be designed to operate above one kelvin~\cite{yang2020operation, petit2022design}.

In these systems, single-qubit operations are commonly implemented by modulating the local electric confinement potential to oscillate the electron across a magnetic field gradient generated by micromagnets fabricated on top of the device. The resulting effective AC magnetic field rotates the spin in a process known as electric-dipole spin resonance (EDSR)~\cite{pioro2008electrically, kawakami2016gate, croot2020flopping}. On the other hand, two-qubit gates can be implemented by taking advantage of the Heisenberg exchange coupling between neighboring QDs~\cite{loss1998quantum, zajac2018resonantly, huang2019fidelity, noiri2022fast, watson2018programmable, petta2005coherent, nowack2011single, kandel2019coherent, takeda2020resonantly, sigillito2019coherent, xue2022quantum, takeda2022quantum, mills2022two, watson2018programmable, veldhorst2015two, xue2019benchmarking}. This coupling can be turned on/off on nanosecond timescales by applying voltage pulses to either lower/raise the inter-dot barrier or by tilting the double well potential. 

QD-based processors containing up to six fully-operational spin qubits have been realized~\cite{philips2022universal}. A promising approach to achieving processors with more qubits is to exploit modularity~\cite{ vandersypen2017interfacing,jnane2022multicore}. For example, architectures in which multiple few-qubit modules are connected by quantum interconnects such as superconducting resonators are particularly attractive for mitigating cross-talk and wiring issues. This combined with techniques such as coherent spin shuttling~\cite{Mills_NatCommun_2019,noiri2022shuttling} provides a route to reach larger QD processors.

Superconducting resonators are widely utilized for mediating long-range interactions between both superconducting qubits~\cite{sillanpaa2007coherent, majer2007coupling, egger2013optimized} and electronic spin qubits~\cite{borjans2020resonant, harvey2022coherent}. While magnetic dipole interactions between electron spins and resonator photons are only in the range of $\mathrm{<kHz}$, strong spin-photon couplings in the range of tens of $\mathrm{MHz}$ may be reached by utilizing the electric dipole interaction in combination with spin-orbit couplings and EDSR techniques~\cite{hu2012strong, benito2017input,mi2017strong, samkharadze2018strong, mi2018coherent, landig2018coherent}. Importantly, by employing microwave photons as mediators, signatures of coherent remote spin-spin interactions have been observed in the resonant~\cite{borjans2020resonant} and dispersive regime~\cite{harvey2022coherent} in QD systems, bringing us a step closer to multicore operations in these processors.

While all-electrical control enables fast gate operations, it has the unwanted side effect of exposing the qubits to incoherent charge noise, which is likely due to charge traps at interfaces in the semiconductor heterostructure ~\cite{yoneda2018quantum, kawakami2016gate, Connors_PRB2019,connors2022charge}. These are dominant sources of decoherence in these systems and are important factors in designing processor architectures. Magnetic noise caused by hyperfine interactions with $^{29} \mathrm{Si}$ nuclear spins can also be an issue. However, this can be largely mitigated by isotopic purification to reach low levels ($50-800~\mathrm{ppm}$)~\cite{xue2022quantum, weinstein2023universal, tyryshkin2012electron,yoneda2018quantum, tanttu2023stability} of residual $^{29} \mathrm{Si}$ isotope concentrations, resulting in coherence times of up to a few milliseconds for spins in gate-defined QDs~\cite{yoneda2018quantum}, and coherence times of up to 10 seconds for the spins of donor-bound electrons~\cite{ tyryshkin2012electron}. 

Despite these advances, theoretical studies of remote spin-spin coupling in semiconductor platforms have so far been limited to two qubits, with one qubit in each module~\cite{srinivasa2016entangling, warren2019long, benito2019optimized, warren2021robust, young2022optimal}. One needs to consider at least three qubits in order to investigate important issues such as the impact of spectator qubits on the performance of resonator-mediated gates. Another important question pertains to the quality of remote entangling operations between qubits that are not directly coupled to the resonator. It is also important to understand how charge noise spreads in modular, multi-qubit systems and how its impact compares to other loss mechanisms, such as cross-talk and leakage.

In this paper, we analyze short-range and long-range entangling operations in a QD system with three spin qubits, in which one qubit is confined to a double quantum dot (DQD) that is resonator-coupled to a triple quantum dot (TQD) module containing two qubits. We examine the performance of short-range (intra-module) and long-range (inter-module) two-qubit entangling gates and explore the effect of spectator qubits. We also study the effect of charge noise on long-distance entangling gate fidelities in the presence of a spectator qubit, finding that the latter has a stronger impact on fidelities for typical experimental parameter regimes. Our work serves as a natural first step towards identifying optimal operating regimes for modular architectures containing multiple qubits connected by resonators.

The paper is organized as follows. In Section~\ref{sec-l1:Three-q-system} we introduce the structure of the resonator-mediated three-qubit system and outline the Hamiltonian, relevant parameters, and the notation. We then present an effective model that accurately captures the dynamics of the dressed spin system which is used to encode the qubits. In Section~\ref{sec-l1:SR-entanglements}, a protocol for implementing short-range entangling gates between two neighboring spin qubits is studied. Long-range entangling gates across the resonator and the effect of the quasistatic charge noise are presented in Section~\ref{sec-l1:LR-entanglements}. Section~\ref{sec-l1:Conclusion} summarizes the results of the study.


\section{\label{sec-l1:Three-q-system}Three-qubit system}
\subsection{\label{sec-l2:Structure}Setup}
\begin{figure*}[t]
\centering
\begin{subfigure}{}
    \centering \includegraphics[width=0.3\textwidth]{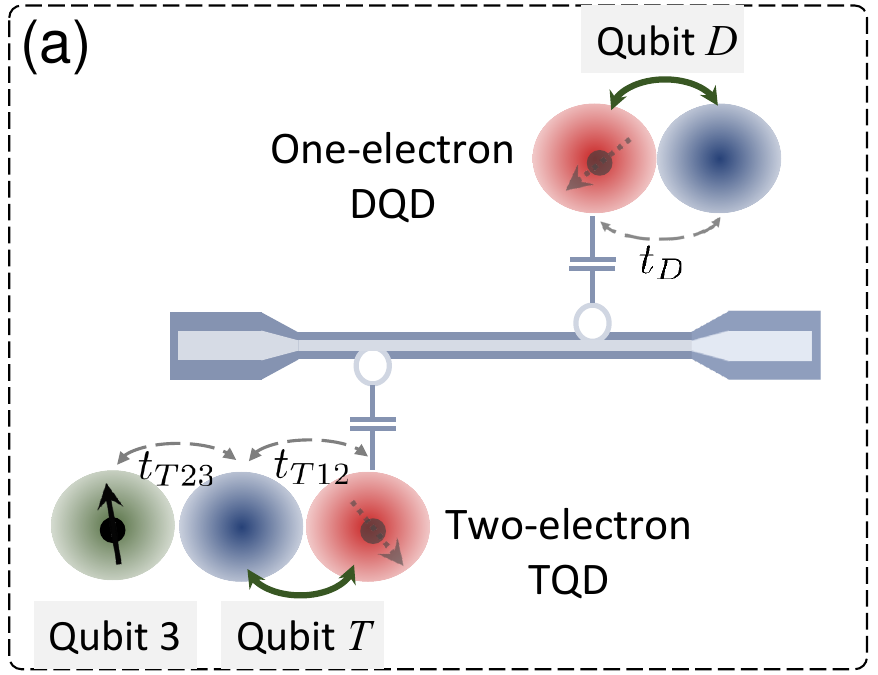}
    \label{fig:1-a}
\end{subfigure}
\begin{subfigure}{}
    \centering \includegraphics[width=0.3\textwidth]{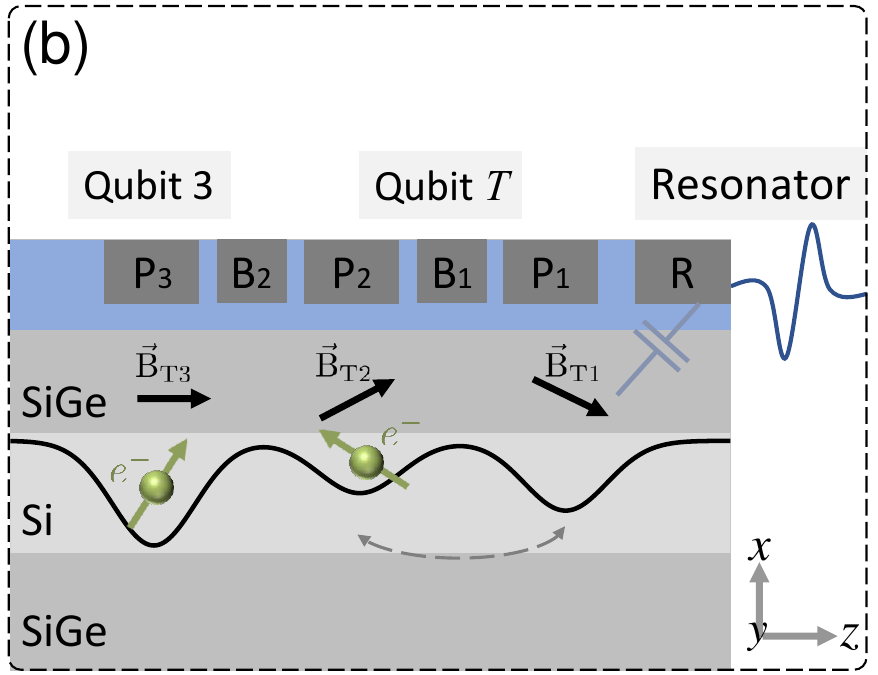}
    \label{fig:1-b}
\end{subfigure}
\caption{(a) An abstract sketch of the three-qubit QD system composed of three main components: a two-electron TQD module, a single-electron DQD module, and a superconducting resonator as a quantum bus with virtual photons to mediate interactions. (b) Illustration of a two-electron TQD module capacitively coupled to a superconducting resonator based on gate-defined QDs in a Si/SiGe heterostructure. The plunger ($\text{P}_1$, $\text{P}_2$, and $\text{P}_3$) and barrier ($\text{B}_1$ and $\text{B}_2$) gates on top of the structure control the chemical potentials of the wells and the inter-dot tunnel couplings. Resonator electrode (R) capacitively couples dot 1 to a probe inside a superconducting resonator. A nearby micromagnet (not shown) creates a static magnetic field gradient ($\vec{B}$) between the first two dots with the arrangement of the magnetic field as illustrated. The depths of the potential wells are not shown to scale for illustrative purposes. The single-electron DQD module has a layout similar to the first two dots of the TQD module.}
\label{fig:1}
\end{figure*}

We consider three electron spin qubits confined to gate-defined QDs. These QDs are created by applying DC voltages on gate electrodes located on the surface of a semiconductor heterostructure to isolate and confine three electrons from the 2D electron gas that resides at an interface within the heterostructure. We consider a device containing a total of five QDs separated into two modules, one of which is a DQD containing one electron and the other a TQD with two electrons. The two modules are connected by a superconducting transmission line resonator, which capacitively couples to one QD from each module. Figure~\ref{fig:1}(a) shows a schematic representation of the hybrid Si-cQED device under study with the three spin qubits in the system labeled. A more detailed illustration of the TQD module in this system is shown in Fig.~\ref{fig:1}(b). The DQD module, also connected to the resonator, has a configuration similar to the first two dots of the TQD module.

The electrical voltage signals applied to the plunger and barrier gate electrodes on top of the device control the electrostatic environments within the heterostructure and provide effective knobs to control the chemical potential level of all dots, $\epsilon_{Di}$ and $\epsilon_{Ti}$, and the inter-dot tunnel couplings, $t_D$ and $t_{Tij}$. Here, we label the two dots in the DQD by $D1$ and $D2$, while the three dots in the TQD are labeled $T1$, $T2$, and $T3$. The resonator couples to dots $D1$ and $T1$.

We assume that the device is functioning in a three-electron regime such that with proper voltage controls the charge occupations are such that there is a single electron delocalized across the DQD (qubit $D$), a single electron delocalized across the first two dots of the TQD (qubit $T$), and one electron trapped in the third dot ($T3$) of the TQD module (qubit $3$).

\subsection{\label{sec-l2:Hamiltonian}Hamiltonian}
The Hamiltonian of the entire system can be represented by modeling the DQD module, TQD module, resonator, drive, and the interaction between the coupled parts, assuming $\hbar=1$, as
\begin{equation}\label{eq:H-Hubbard-1}
    \begin{aligned}
            \hat{H}_{\text{tot}} = {} & \hat{H}_{r}+\hat{H}_{\text{DQD}}+\hat{H}_{\text{TQD}}+\hat{H}_{\text{DQD-R}}\\
            & +\hat{H}_{\text{TQD-R}}+\hat{H}_{\text{drive}}, \\
            \hat{H}_{r} = {} & \omega_{r} a^{\dagger} a, \\
            \hat{H}_{\text{DQD}} = {} & \sum_{i=1}^{2}\left(\epsilon_{Di} n_{Di}+\vec{B}_{Di} \cdot \vec{S}_{Di}\right)  \\
            & -\sum_{\sigma=\uparrow, \downarrow}\left(t_{D} ~c_{D2\sigma}^{\dagger} c_{D1\sigma}+\text{H.c.}\right), \\
            \hat{H}_{\text{TQD}} = {} & \sum_{i=1}^{3}\left(\epsilon_{Ti} n_{Ti}+\frac{U_{Ti}}{2}~ n_{Ti}\left(n_{Ti}-1\right)+\vec{B}_{Ti} \cdot \vec{S}_{Ti}\right)\\
            & +V_{T12}~n_{T1} n_{T2}+V_{T23}~n_{T2} n_{T3}+V_{T31}~n_{T3} n_{T1} \\
            & -\sum_{\sigma=\uparrow, \downarrow}\Big( \Big. t_{T12}~c_{T2\sigma}^{\dagger} c_{T1\sigma}+t_{T23}~c_{T3\sigma}^{\dagger} c_{T2s}\\
            & +t_{T31}~c_{T1\sigma}^{\dagger} c_{T3\sigma}+\text {H.c.}\Big. \Big), \\
             \hat{H}_{\text{DQD-R}} = {} & 2g_{D}^{A C}\left(a^{\dagger}+a\right)n_{D1},\\
             \hat{H}_{\text{TQD-R}} = {} & 2g_{T}^{A C}\left(a^{\dagger}+a\right)n_{T1},\\
             \hat{H}_{\text{drive}} = {} &
             \Omega_D\cos\left(\omega_dt\right)\left(n_{D1}-n_{D2}\right),
     \end{aligned}
\end{equation}
where $\hat{H}_{r}$ describes the microwave resonator, while the $\hat{H}_{\text{DQD}}$ and $\hat{H}_{\text{TQD}}$ terms describe the DQD and TQD modules. The Hamiltonians for the DQD and TQD modules correspond to a single-band Fermi-Hubbard model. This Hamiltonian has been previously examined in the case of semiconductor quantum dot spin qubits~\cite{yang2011generic, sarma2011hubbard}. The capacitive couplings between the resonator field mode and the charge occupation of the nearest dot to the resonator are modeled by the terms $\hat{H}_{\text{DQD-R}}$ and $\hat{H}_{\text{TQD-R}}$. 

Here $a^{\dagger}$($a$) is the bosonic creation (annihilation) operator of the resonator mode with the angular frequency $\omega_{r}$ and $c_{i\sigma}^{\dagger}$($c_{i\sigma}$) are the fermionic creation (annihilation) operators for an electron in dot $i$ with spin $\sigma$. Operator $n_k$ is the total number operator for dot $k$, i.e., $n_k=n_{k\uparrow}+n_{k\downarrow}$, with $n_{k\uparrow}$ ($n_{k\downarrow}$) being the number operator for the spin up (down) state in the dot $k$. The spin vector for dot $k$ is $\vec{S}_{k}=\left(c_{k\uparrow}^{\dagger} c_{k\downarrow}+c_{k\downarrow}^{\dagger} c_{k\uparrow},\frac{c_{k\uparrow}^{\dagger} c_{k\downarrow}-c_{k\downarrow}^{\dagger} c_{k\uparrow}}{i},c_{k\uparrow}^{\dagger} c_{k\uparrow}-c_{k\downarrow}^{\dagger} c_{k\downarrow}\right)$. We assume that only one bosonic mode of the resonator is relevant to the dynamics of the system. We also only include the ground states of the confinement potentials of the QDs in the orbital part of the Hilbert space since the excited orbital states are at sufficiently higher energies. The chemical potentials of the dots are denoted by $\epsilon_{Di}$ and $\epsilon_{Ti}$, and the inter-dot tunnel couplings are captured by the $t_D$ and $t_{Tij}$ parameters, determined by the spatial overlap of the wave functions in the dots. In practice, both sets of parameters can be electrically tuned using gate electrodes. $U$ and $V$ are the intra- and inter-dot Coulomb repulsion energies, respectively, and they are found to be typically in the range of a few to ten meV~\cite{yang2011generic,sarma2011hubbard}. To impose the condition that one electron in the TQD module gets localized to dot $T3$ with high probability, we impose $\epsilon_{T3} \ll \epsilon_{T2}+V_{T12}-V_{T31},\epsilon_{T1}+V_{T12}-V_{T23} \ll U_{T1}, U_{T2}, U_{T3}$ in all the analyses that follow.

While the effect of intrinsic spin-orbit interactions can be ignored, artificial spin-orbit interactions are engineered by placing micromagnets on top of the device and creating magnetic field gradients across neighboring dots. Leveraging this technique, the interaction between the spin of delocalized electrons and the electric field from the resonator is mediated through electric dipole interactions~\cite{hu2012strong}, leading to higher effective spin-photon coupling strengths~\cite{mi2017strong, samkharadze2018strong, mi2018coherent, landig2018coherent} and therefore practical two-qubit interaction times. The electric dipole coupling strengths for coupling to the resonator fields are denoted by $g_D^{AC}$ and $g_T^{AC}$ for the DQD and TQD modules, respectively. The total magnetic field vector, including the external field as well as that generated by the micromagnets, in dot $i$ of DQD and TQD is denoted by $\vec{B}_{Di}$ and $\vec{B}_{Ti}$. In the numerical studies that follow, the average magnetic field in the DQD and the first two dots of the TQD are chosen along the $z$ direction, while the magnetic field gradient is along the $x$ axis to induce transverse spin-orbit couplings. The magnetic field in the third dot of the TQD is along the $z$ direction (Fig.~\ref{fig:1}(b)).

The Hamiltonian of Eq.~(\ref{eq:H-Hubbard-1}) also includes an electric drive term on the potentials of the dots in the DQD module with amplitude $\tilde{\Omega}$ and frequency $\omega_d$. This drive is used in the implementation of a long-range maximally entangling gate (in this case CNOT), as discussed in Section~\ref{sec-l1:LR-entanglements}.

For the one- and two-electron charge configurations assumed above for the DQD and TQD modules, respectively, the single-particle subspace of the DQD has a dimensionality of four, with computational basis states given by $c_{D1 \uparrow}^{\dagger}\ket{0}$, $c_{D1 \downarrow}^{\dagger}\ket{0}$, $c_{D2 \uparrow}^{\dagger}\ket{0}$, $c_{D2 \downarrow}^{\dagger}\ket{0}$, while the two-particle subspace of the TQD has a dimensionality of fifteen, with computational basis states given by $c_{T1 \uparrow}^{\dagger} c_{T3 \uparrow}^{\dagger}\ket{0}$, $c_{T1 \uparrow}^{\dagger} c_{T3 \downarrow}^{\dagger}\ket{0}$, $c_{T1 \downarrow}^{\dagger} c_{T3 \uparrow}^{\dagger}\ket{0}$, $c_{T1 \downarrow}^{\dagger} c_{T3 \downarrow}^{\dagger}\ket{0}$, $c_{T2 \uparrow}^{\dagger} c_{T3 \uparrow}^{\dagger}\ket{0},c_{T2 \uparrow}^{\dagger} c_{T3 \downarrow}^{\dagger}\ket{0}$, $c_{T2 \downarrow}^{\dagger} c_{T3 \uparrow}^{\dagger}\ket{0}$, $c_{T2 \downarrow}^{\dagger} c_{T3 \downarrow}^{\dagger}\ket{0}$, $c_{T1 \uparrow}^{\dagger} c_{T2 \uparrow}^{\dagger}\ket{0},c_{T1 \uparrow}^{\dagger} c_{T2 \downarrow}^{\dagger}\ket{0}$, $c_{T1 \downarrow}^{\dagger} c_{T2 \uparrow}^{\dagger}\ket{0}$, $c_{T1 \downarrow}^{\dagger} c_{T2 \downarrow}^{\dagger}\ket{0}$, $c_{T1 \uparrow}^{\dagger} c_{T1 \downarrow}^{\dagger}\ket{0}$, $c_{T2 \uparrow}^{\dagger} c_{T2 \downarrow}^{\dagger}\ket{0}$, $c_{T3 \uparrow}^{\dagger} c_{T3 \downarrow}^{\dagger}\ket{0}$.
It should be noted that although the multi-qubit dynamics are purposely designed to primarily happen in the low-energy subspace of the full system Hamiltonian, all the simulations are carried out over the entire Hilbert space of Eq.~\eqref{eq:H-Hubbard-1}, after we truncate the maximum resonator photon number to $n_r$. This results in a total system Hilbert space dimension of $60n_r$ that we use in our numerical simulations below.

In this work, we focus on TQDs arranged in a linear array, i.e., the three QDs sit on a line (see Fig. ~\ref{fig:1}(a)), with only nearest-neighbor tunnel coupling, i.e., $t_{T13}=0$. This scheme is relevant to physical implementations of gate-defined QD-based quantum processors that have been the subject of investigations recently~\cite{philips2022universal}.
\subsection{\label{sec-l2:Testing the effective model}Testing the effective model}

We consider the case of a dispersive regime, where coupling terms are sufficiently small relative to energy differences, and the drives are weak and detuned and can be considered perturbatively in the analysis. By applying frame transformations and a time-dependent Schrieffer-Wolff transformation~\cite{schrieffer1966relation}, the coupling between the low-energy subspace and the rest of the Hilbert space is removed to the leading order (see Appendix \ref{appendix:A-Eff-Ham-derivation} for the details of the derivation). Next, we project onto the low-energy subspace defined as the empty-cavity limit, i.e., $\braket{a^\dagger a}=0$, and the ground state for the orbital degree of freedom of the DQDs, i.e., $\forall i \in \{D,T\}\colon \braket{\tau_i^z}=-1$, through which we derive an effective Hamiltonian, without the drive, governing the dynamics of the low-energy qubits:
\begin{equation}\label{eq:H-Eff-Maintext}
            \hat{H}_{\mathrm{eff}}=\sum_{i=D, T, 3} \frac{1}{2} \omega_i \sigma_i^z-J_r \sigma_D^x \sigma_T^x+\frac{J_e}{4} \vec{\sigma}_T \cdot \vec{\sigma}_3+J_{Z Z} \sigma_T^z \sigma_3^z.
\end{equation}
Here the effective Hamiltonian is composed of six terms. The first three terms represent the three dressed qubits with transition frequencies $\omega_i,~i \in \{D,T,3\}$. The fourth term describes the long-distance resonator-mediated coupling between qubits $D$ and $T$ with strength $J_r$. The fifth term corresponds to a short-distance exchange coupling between qubits $T$ and $3$ with strength $J_e$. The last term is a residual short-distance Ising coupling with strength $J_{ZZ}$ between qubits $T$ and $3$ resulting from spin-charge hybridization. For the parameter regimes considered in this paper, this residual coupling is small and can be safely ignored if desired. A key observation here is that the structure of the effective Hamiltonian is comprised of two bipartite interactions, each between two neighboring qubits, as one might anticipate. However, the coupling strengths of these interactions depend on the details of the entire system (see Appendix \ref{appendix:A-Eff-Ham-derivation} for the definitions of the effective couplings).

By engineering the dynamics in an unpopulated resonator mode, long-distance interactions are mediated via the exchange of virtual photons. The benefits of choosing such a subspace are twofold. First, the cavity-induced losses are effectively removed; and second, operations through the quantum bus between different sets/groups of modules can be parallelized, as interactions are merely virtual and no real populations are induced in the resonator, thus increasing the information processing speed and the throughput of the system.

Next, we put the derived effective Hamiltonian to the test through a numerical fitting approach, as described below. Confirming the validity of the effective Hamiltonian is important since we will use it extensively later on to devise protocols for implementing entangling operations between pairs of qubits.

\begin{figure*}[t]
\centering
\begin{subfigure}{}
    \centering \includegraphics[width=0.32\textwidth]{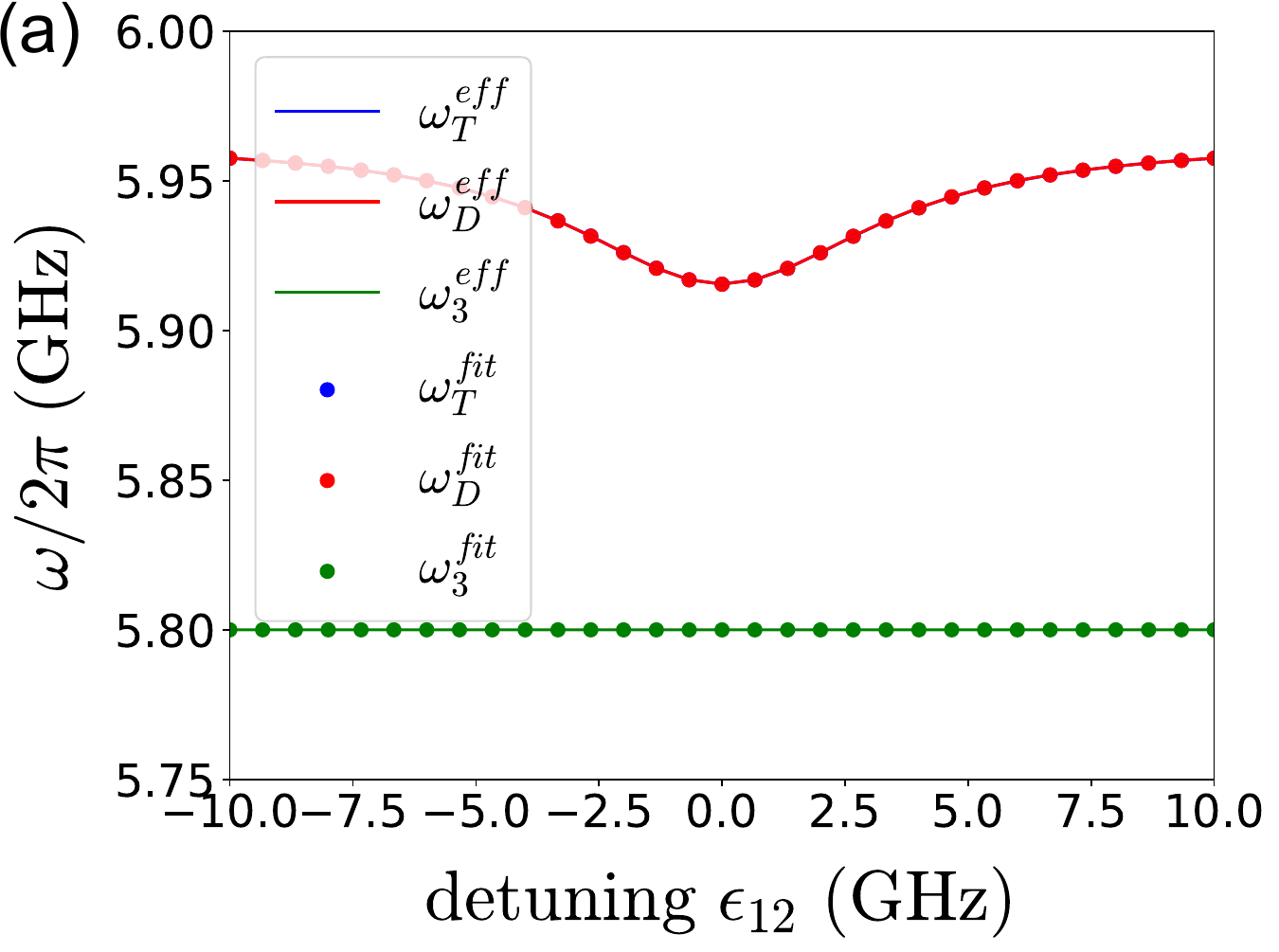}
    \label{fig:2-a}
\end{subfigure}
\begin{subfigure}{}
    \centering \includegraphics[width=0.32\textwidth]{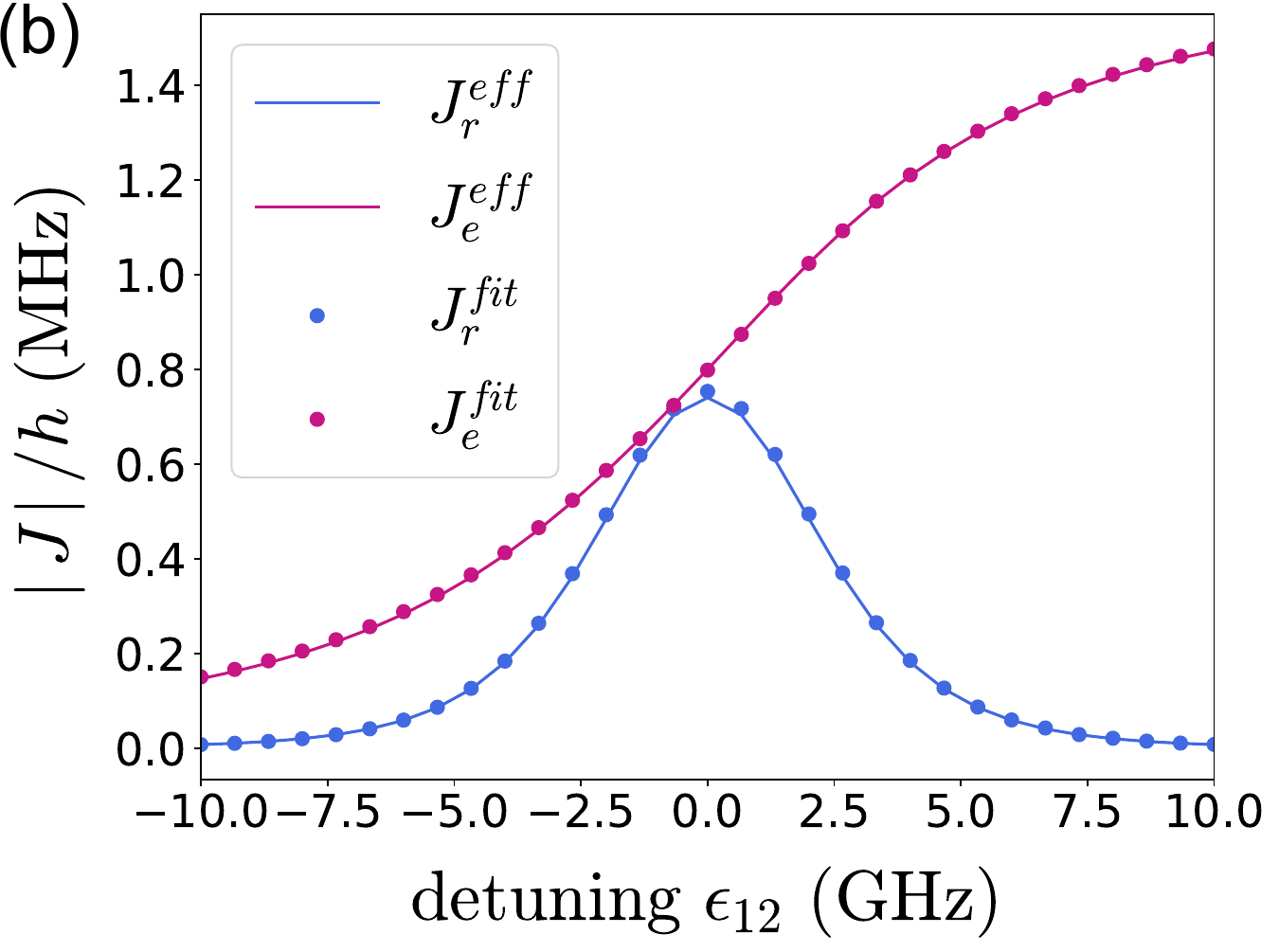}
    \label{fig:2-b}
\end{subfigure}
\begin{subfigure}{}
    \centering \includegraphics[width=0.32\textwidth]{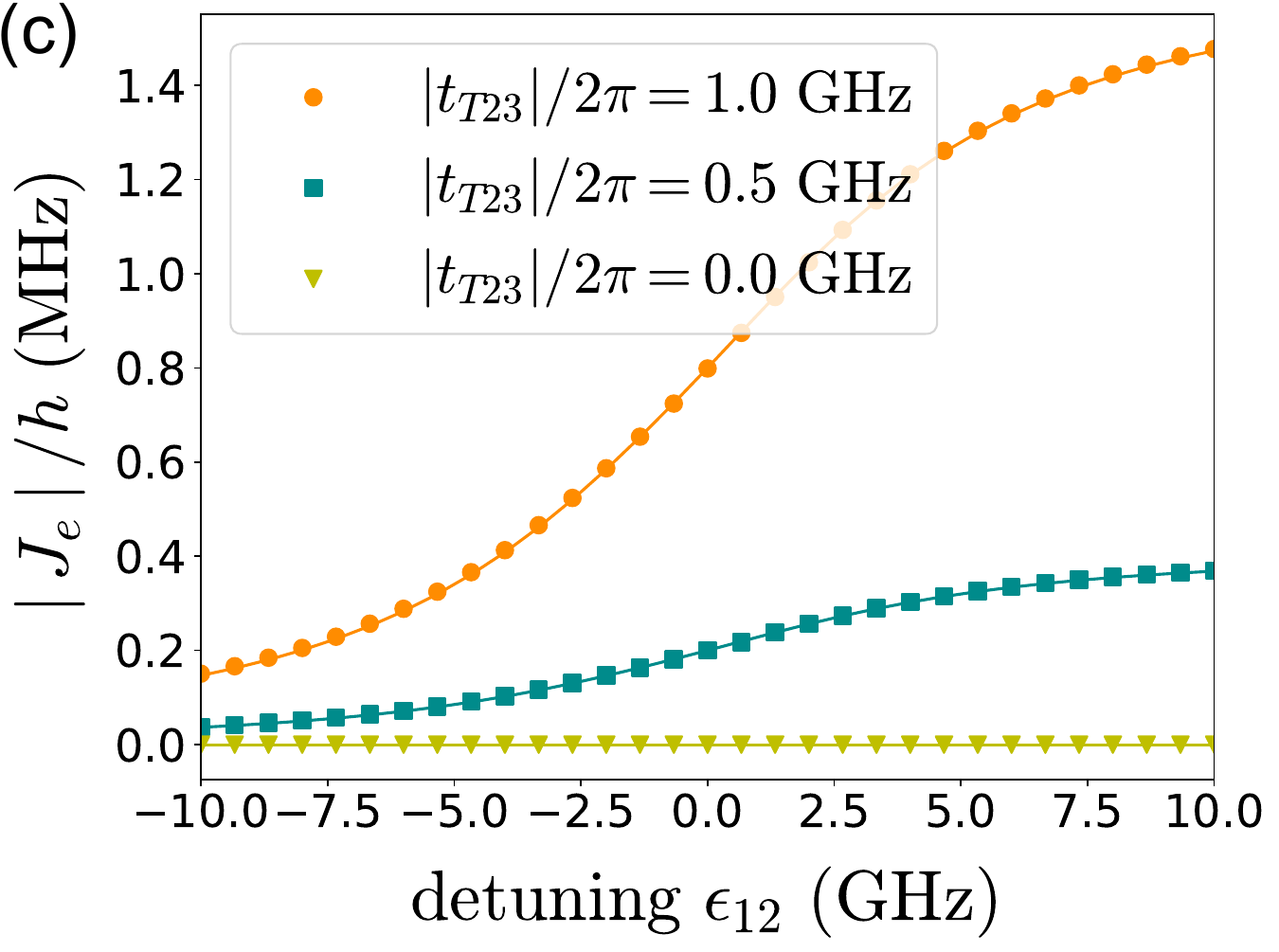}
    \label{fig:2-c}
\end{subfigure}
\caption{Numerically derived values of the effective Hamiltonian parameters (dots) and the same parameters calculated from the analytically derived effective model (lines) for the dressed qubits’ transition frequencies (a), resonator-mediated coupling strength (b), and exchange-enabled coupling strengths (b,c) vs the detuning, $\epsilon$, of the DQD and the first two dots of the TQD module. Parameters are set at $\omega_r/2\pi=6~\mathrm{GHz}, \omega_T^z/2\pi=\omega_D^z/2\pi=5.95~\mathrm{GHz}$ ($z$-directed magnetic field strength in the DQD and the first two dots of the TQD), $\omega_3^z/2\pi=5.8~\mathrm{GHz}$ ($z$-directed magnetic field strength in the third dot of TQD, with no $x$-component), $t_D/2\pi=t_{T 12}/2\pi=3.5~\mathrm{GHz}, t_{T 23}/2\pi=1~\mathrm{GHz}$ (a,b), $ t_{T 23}/2\pi=0,~0.5,~1~\mathrm{GHz}$ (c), $g^x_T/2\pi=g^x_D/2\pi=200~\mathrm{MHz}, g_D^{AC}/2\pi=g_T^{AC}/2\pi=40~\mathrm{MHz}, \epsilon_{T3}/2\pi=-300~\mathrm{GHz}, U_{T1}/2\pi= U_{T2}/2\pi= U_{T3}/2\pi=2.5~\mathrm{THz}$.}
\label{fig:2}
\end{figure*}

We start by finding the time evolution of the whole system, $U_{\text{tot}} (t)= e^{-i t \hat{H}_{\text{tot}}}$ for $t\in[0,t_0]$, followed by projection onto the low-energy subspace with the projection operator $P$, which yields the actual time evolution of the low-energy subspace $U_{\text{actual}}(t)=P U_{\text{tot}}(t)P$. The details of the projection operation and low-energy subspace are introduced in Appendix \ref{appendix:A-Eff-Ham-derivation}. We then compare the actual time evolution with its counterpart based on the effective model, $U_{\text{eff}}(T)=e^{-i t H_{\text{eff}}}$. The metric for this comparison is the transformation fidelity defined as $ F=\frac{dF_e+1}{d+1}$, where $ F_e=\frac{1}{d^2}{ \lvert \text{Tr}[{U_{\text{actual}}^\dagger(t) U_{\text{eff}}(t)}] \rvert}^2$ is the process fidelity \cite{pedersen2007fidelity}, and $d$ is the subspace dimension. We optimize this fidelity with respect to all the parameters of the effective model to find the best effective parameters (Fig.~\ref{fig:2}). Optimizing for only one point in the evolution time results in an infinite number of optimal parameter values, all giving the same fidelity, due to a periodic dependence on the parameters. To resolve this issue, we instead optimize the average fidelity over a sample of points during a time interval. The starting values in these optimizations are the analytically-derived effective parameters (see Appendix \ref{appendix:A-Eff-Ham-derivation}) and numerical simulations are truncated at three resonator levels. A time duration of $t_0=20~\mu$s is used for the optimization procedure.

Figure~\ref{fig:2} shows the transition frequencies of the dressed qubits and the resonator-mediated and exchange interaction strengths as a function of the detuning (i.e., the tilt of a double well potential), $\epsilon$, of the DQD and the first two dots of the TQD module. The dots in the figure correspond to numerically found quantities from the fitting scheme explained above, and the lines correspond to the same parameters obtained from the effective model. The parameters used in the simulation are reported in the figure caption.
The fidelity of the fit is greater than $97.5\%$, with a small residual coupling strength of $\lvert J_{ZZ} \rvert/h < 4~\mathrm{kHz}$.

The fact that for all quantities the numerically found effective values are nearly identical to the matching values from the analytical model confirms that the effective model accurately captures the behavior of the low-energy three-qubit computational space and can be safely used to understand the dynamics of the system and devise gate protocols. Next, we study gate designs for short-range and long-range entangling gates, in the regime where the effect of the spectator qubit is suppressed, and we examine the interference caused by the spectator qubit on the fidelity of two-qubit operations. The simulation parameters employed throughout this work are chosen consistent with values from recent experimental works~\cite{mi2018coherent, xue2022quantum, noiri2022shuttling, mills2022two}.


\section{\label{sec-l1:SR-entanglements}Short-range entangling gates}

In this section, we describe and compare several protocols for implementing short-range entangling gates between qubits $T$ and $3$ within the TQD module. We focus on the implementation of CPHASE gates, which in combination with single-qubit gates yields a universal quantum gate set.

The effective two-qubit interaction between the qubits in the TQD module (Eq.~\eqref{eq:H-Eff-Maintext}), in the regime where the effect of the third qubit is suppressed, is essentially in the form of an extended Heisenberg interaction between the two dressed spins residing in the TQD module:
\begin{equation}\label{eq:H-T3-Heisenberg}
            \hat{H}_{\mathrm{eff},~J_r \to 0}=\frac{1}{2} \omega_D \sigma_D^z+\frac{1}{2} \omega_T \sigma_T^z+\frac{1}{2} \omega_3 \sigma_3^z +\frac{J_e}{4} \vec{\sigma}_T \cdot \vec{\sigma}_3.
\end{equation}
In view of this, in principle, different techniques may be employed to realize two-qubit operations, such as AC-pulsed frequency-selective CROT and CNOT gates~\cite{zajac2018resonantly, huang2019fidelity, noiri2022fast, watson2018programmable}, DC-pulsed~\cite{petta2005coherent, nowack2011single, kandel2019coherent} and resonantly driven SWAP~\cite{takeda2020resonantly, sigillito2019coherent} gates, and CPHASE gates~\cite{xue2022quantum, takeda2022quantum, mills2022two, watson2018programmable, veldhorst2015two, xue2019benchmarking}.

Two-qubit gate implementations primarily depend on the strength of the exchange coupling, $J_e$, relative to the difference in the energy splitting of the two qubits, $\Delta\omega=\lvert \omega_T-\omega_3 \rvert$. For $J_e \gg \Delta\omega$, the exchange interactions result in SWAP evolutions, as the eigenstates of the two-qubit space are the spin singlet and triplets, which was in fact the original proposal to achieve entangling gates in QDs~\cite{loss1998quantum}. This regime is accessible by making the DQD energy levels highly detuned (to approximately $10-15~\mathrm{GHz}$ range~\cite{petta2005coherent}), reaching a few~$\mathrm {GHz}$ strength~\cite{petta2005coherent, nowack2011single, kandel2019coherent} for the inter-dot exchange couplings, making SWAP operations accessible.

Considering that charge noise is one of the important sources of incoherent noise in these electrically controlled systems ~\cite{yoneda2018quantum, kawakami2016gate, eng2015isotopically, benito2019optimized}, it is crucial to make gate protocols that can be applied in the regimes where the impact of this noise source is small. When a DQD is tuned to the symmetric operation point (zero detuning), the effect of the charge noise is suppressed to first order~\cite{reed2016reduced, bertrand2015quantum, martins2016noise}, resulting in longer coherence times. However, only smaller values of $J_e$ are typically accessible at these points, and to get larger $J_e$ the operation point needs to be moved away from the symmetric point, causing more decoherence in the system~\cite{petta2005coherent}. Besides, many QD spin qubit systems utilize large magnetic field gradients, enabling the addressability of the qubits. In practice exchange coupling strengths are in the range of hundreds of $\mathrm{kHz}$ to $\mathrm{10-20~MHz}$~\cite{noiri2022shuttling, xue2022quantum}, which is much smaller than the difference in the transition frequencies of the neighboring qubits, which is typically hundreds of $\mathrm{MHz}$ to $\mathrm{GHz}$~\cite{xue2022quantum, noiri2022shuttling}.

For these reasons, here we focus on the $J_e \ll \Delta\omega$ regime, which is naturally relevant for gate-defined QDs with micromagnet-induced magnetic field gradients and tunnel barrier-controlled exchange couplings. Evolution under an exchange interaction in this regime leads to a CPHASE gate between the two qubits~\cite{meunier2011efficient, russ2018high}. CROT gates can also be achieved in this regime by selectively driving one of the EDSR transitions, in order to make a universal gate set. We investigate fast adiabatic CZ gates between qubits $T$ and $3$, which may be implemented by applying a pulse to the barrier gate between dots $2$ and $3$ of the TQD module to temporarily modify the exchange coupling, while qubit $D$ evolves in isolation.

The exchange-based CZ gate can be realized due to the fact that a nonzero exchange interaction $J_e$, in the regime of $J_e \ll \Delta\omega$, lowers the energy of states with antiparallel spins, $\ket{\uparrow\downarrow},\ket{\downarrow\uparrow}$, i.e., the odd-parity subspace, compared to the case with $J_e=0$. This energy shift of antiparallel spins makes it possible to add spin-dependent phase shifts to the odd-parity subspace and thus realize CZ gates in the equivalent evolution form of $ \mathrm{diag}(1,\exp\left({i \phi_{\uparrow\downarrow}}\right),\exp\left({i \phi_{\downarrow\uparrow}}\right),1)$ ~\cite{meunier2011efficient, russ2018high}. Here $\phi_{\uparrow\downarrow}(\phi_{\downarrow\uparrow})=\delta E_{{\uparrow\downarrow}({\downarrow\uparrow})}t_{\mathrm{wait}} $, where $ t_{\mathrm{wait}}$ is the evolution time, while $\delta E_{{\uparrow\downarrow}({\downarrow\uparrow})}$ is the energy shift of the dressed antiparallel eigenstates $\widetilde{\ket{\uparrow\downarrow}}$ and $\widetilde{\ket{\downarrow\uparrow}}$ relative to the states $\ket{\uparrow\uparrow}$ and $\ket{\downarrow\downarrow}$ as a consequence of the exchange interaction. A CZ gate is realized for $\phi_{\uparrow\downarrow}+\phi_{\downarrow\uparrow}=\left(2n+1\right)\pi$ up to local $Z$ gates on qubits one and two ($\exp\left({i\phi_1\sigma_1^z}\right)\exp\left({ i\phi_2\sigma_2^z }\right)$) with phases $\phi_1=-\frac{1}{4}(\int_{0}^{t_g} \sqrt{J_e^2(t)+\Delta\omega^2} \,dt)$ and $\phi_2=\frac{1}{4}(\int_{0}^{t_g} \sqrt{J_e^2(t)+\Delta\omega^2} \,dt)$, respectively. These local $Z$ rotations can be implemented virtually by changing the phases of subsequent pulses and rotating the frame~\cite{mckay2017efficient}. For $J_e \ll \Delta\omega$, the energy shift of the dressed eigenstates, to first order in $J_e$, is equal to $J_e/2$, creating CZ gates at gate time $t_g=\left(2n+1\right)\pi/\bar{J_e}$, where the time-averaged exchange coupling is defined as $\bar{J_e}=\frac{1}{t_g}\int_{0}^{t_g} J_e(t) \,dt$. Additionally, a spin-echo mechanism, adding mid-sequence $\pi$ rotations on both spins about the $x$-axis, can be incorporated to eliminate excess phases during the CZ exchange gate due to noisy unbalanced magnetic fields or local quasi-static phase noise, making the evolution in the even-parity subspace effectively trivial~\cite{russ2018high}.

Although rectangular pulses can be used to generate CZ gates, the intrinsic non-adiabatic nature of such pulses results in coherent errors in the gate dynamics, causing reduced gate fidelities. Since turning on and off the exchange interaction only affects the odd-parity subspace of the effective two-QD system, the full dynamics of the system can be designed by simply engineering the dynamics of the odd-parity subspace spanned by the $\ket{\uparrow\downarrow},\ket{\downarrow\uparrow}$ states, with effective Pauli operators defined as $\sigma_x'=\ket{\uparrow\downarrow}\bra{\downarrow\uparrow}+\ket{\downarrow\uparrow}\bra{\uparrow\downarrow}$ and $\sigma_z'=\ket{\uparrow\downarrow}\bra{\uparrow\downarrow}-\ket{\downarrow\uparrow}\bra{\downarrow\uparrow}$, in terms of which the Hamiltonian in this subspace is
\begin{equation}\label{eq:H-T3-Heisenberg-odd-parity}
            \hat{H}_{\mathrm{odd}}=\frac{1}{2}\left(-J_e(t)+ \Delta\omega \sigma_z’+J_e(t)\sigma_x’\right).
\end{equation}
The requirement for the CZ operation is now a global phase shift in this subspace, while SWAP-like operations due to the non-adiabatic nature of the pulse translate to nontrivial dynamics in the subspace and therefore must be suppressed. The evolution in the odd-parity subspace can be visualized in terms of a Bloch sphere trajectory controlled by the Hamiltonian vector $\left(J_e/2,0, \Delta\omega/2\right)$. In this case, Ref.~\cite{martinis2014fast} showed that the non-adiabatic error, $P_e$, is proportional to the power spectral density of the rate of change of the control Hamiltonian, $d\theta/dt$, with $\theta$ being the angle of the Hamiltonian vector with respect to the quantization axis $z$, calculated at the precession frequency $\omega_q$ as $P_e=\left(1/4\right)S_{d\theta/dt}\left(\omega_q\right)$, in which $S$ denotes the power spectral density. The precession frequency is the difference in the eigenenergies of the two-dimensional subspace, which based on Eq.~\eqref{eq:H-T3-Heisenberg-odd-parity}, is given by $\omega_q=\sqrt{J_e^2+\Delta\omega^2}$.

Now, with the assumption of $ J_e \ll \Delta\omega$, which is relevant to the case studied here, the non-adiabatic error will be relatively small, and the connection between the non-adiabatic error and the power spectral density of the signal holds in this small-error limit. This is due to the large precession frequencies of the Bloch vector (around the Hamiltonian vector) compared to the relatively slower time dependence of the change in $\theta$ over the Bloch sphere. For the Hamiltonian~\eqref{eq:H-T3-Heisenberg-odd-parity}, the direction of the ground eigenstate Bloch vector is antiparallel with the Hamiltonian vector, and therefore the rate of change of the control field is proportional to the rate of change of the instantaneous eigenstates of the system. To have simply a global phase in the odd-parity subspace, the Hamiltonian vector should not deviate much from the $z$-axis, consistent with the condition of $ J_e(t) \ll \Delta\omega$. We specifically study two cases: (i) a rectangular pulse with constant $J_e$ during the gate time, corresponding to a fixed $\theta$ during the pulse, and (ii) smooth $J_e$ signals using window functions, for which $\theta$ slowly varies from zero to a maximum angle and then returns. Both control schemes have been experimentally utilized to generate CZ gates. See Refs.~\cite{mills2022two, watson2018programmable, veldhorst2015two, xue2019benchmarking} for CZ gates with rectangular control signals and Refs.~\cite{xue2022quantum, takeda2022quantum} for CZ gates with Hann signals. 

The rate of non-adiabatic errors can be investigated by moving to the adiabatic frame of the system, using the transformation $T=\exp\left({-i\arctan(J_e(t)/\Delta\omega)\sigma_y'/2}\right)$, or equivalently by analyzing the change in the Hamiltonian angle $d\theta/dt=\dot{J}_e /\Delta\omega$, resulting in non-adiabatic errors $P_e \propto S_{\left(\dot{J}_e/\Delta\omega \right)}\left(\omega_q\right)$. 

With the rectangular CZ pulse, $J_e(t)=J_0/2,~0<t<t_g$, the CZ gate is reached at $ t_g=2\pi/J_0$, and the non-adiabatic error is given by $P_e^{\mathrm{rect}} \propto \frac{1}{\alpha}\sin^2\left(\omega_q t_g/2\right)$ with $\alpha=\left(\Delta\omega t_g/\pi\right)^2$. The smallest gate time at which the non-adiabatic error is minimized is $t_g=2\pi/\omega_q$ (synchronization). However, these pulses have an infinitely fast rise and fall, which is not quite practical due to the cut-off frequencies of the control electronics. Also, the fall-off rate of the non-adiabatic error with increasing gate time is quite slow $\left(\sim1/t_g^2\right)$ for these pulses. Besides, it is crucial to have stable control signals, such that the error does not increase rapidly under small changes in the applied signal. Because of these factors, smooth control pulses that perform in a way that is insensitive to slight variations in the waveform shape are generally of interest.

These issues can be addressed by replacing the rectangular pulse with a smooth pulse that optimizes the energy spectral density. This is a known task in the field of signal processing~\cite{prabhu2014window}, and here we leverage some prior results from that literature to study gates for our three-qubit system. We employ the Hann window function, $W_\mathrm{Hann}(t)=\frac{1}{2}\left(1-\cos(2\pi t/t_g)\right)$, which is an extensively used function in signal processing and is essentially the first-order Fourier window function. For the Hann pulse, $J_e(t)=J_0 W_\mathrm{Hann}$, and the CZ gate is again reached at $t_g=2\pi/J_0$. The non-adiabatic error can be calculated as $P_e^\mathrm{Hann}\propto \frac{1}{\alpha}\sin^2\left(\omega_q t_g/2\right)/\lvert 1-\left(\omega_0 t_g/2\pi\right)^2\rvert^2$ with the same $\alpha$ parameter as $ P_e^\mathrm{rect}$. 

\begin{figure*}[t]
\centering
\begin{subfigure}{}
    \centering \includegraphics[width=0.33\textwidth]{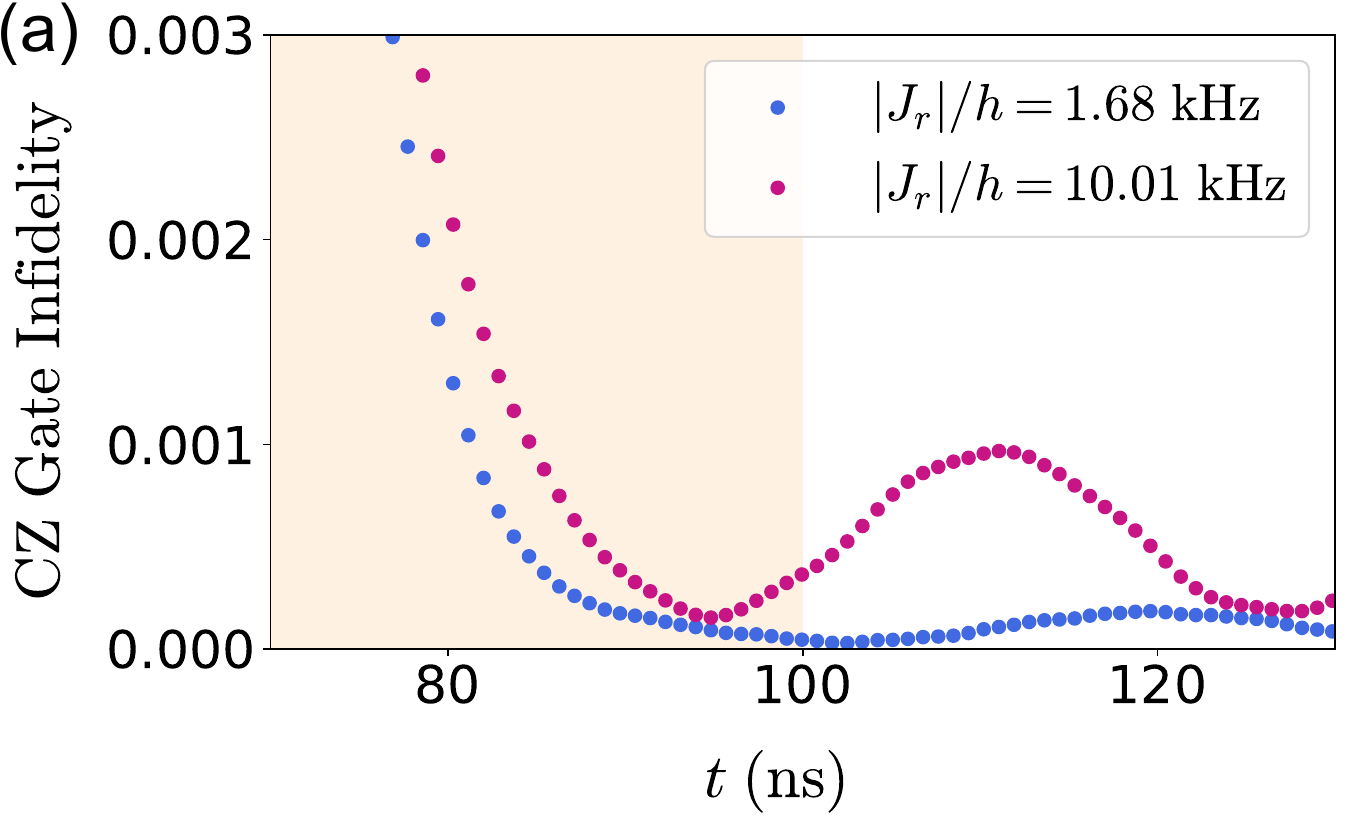}
    \label{fig:3-a}
\end{subfigure}
\begin{subfigure}{}
    \centering \includegraphics[width=0.33\textwidth]{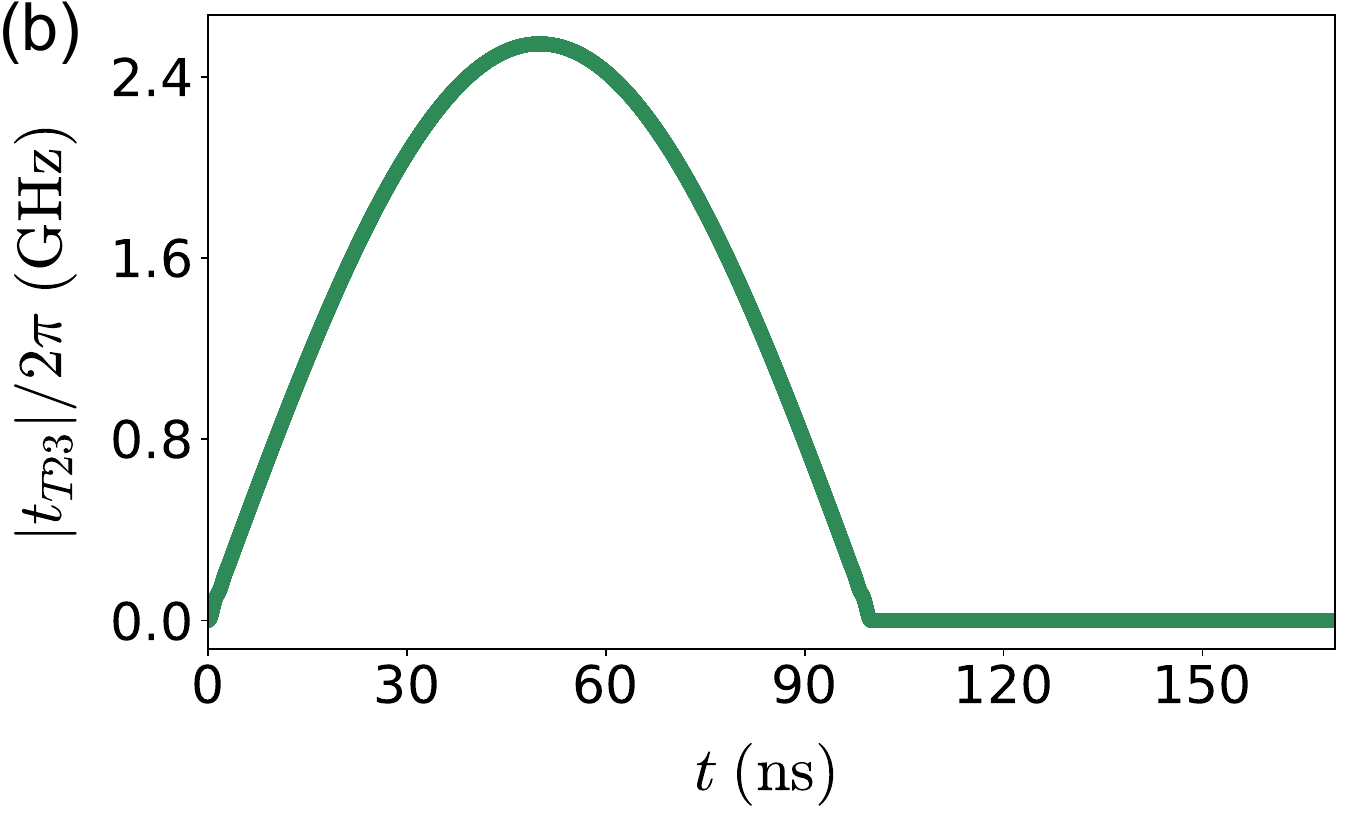}
    \label{fig:3-b}
\end{subfigure}
 \\ 
    \vspace{0.8cm}
\begin{subfigure}{}
    \centering \includegraphics[width=0.33\textwidth]{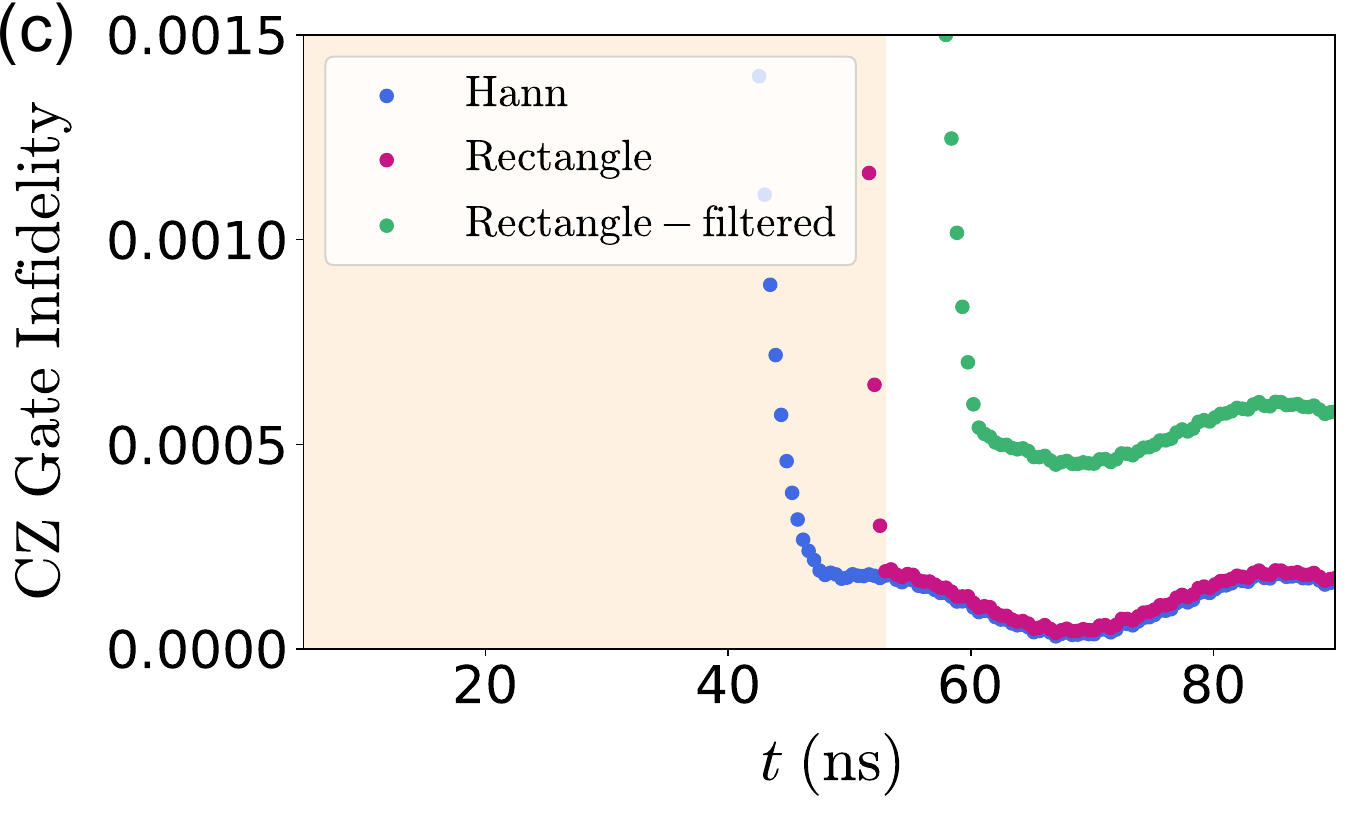}
    \label{fig:3-c}
\end{subfigure}
\begin{subfigure}{}
    \centering \includegraphics[width=0.33\textwidth]{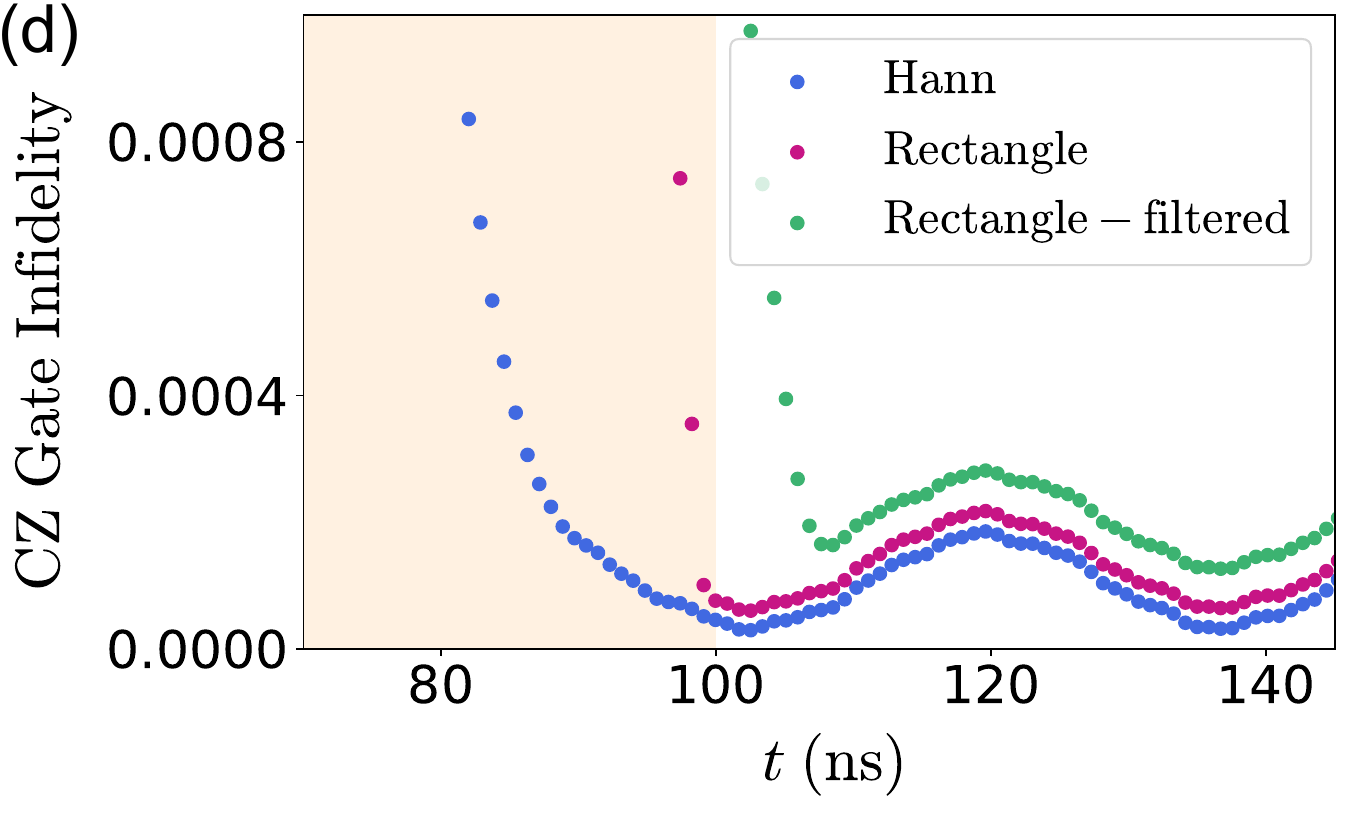}
    \label{fig:3-d}
\end{subfigure}
\begin{subfigure}{}
    \centering \includegraphics[width=0.3\textwidth]{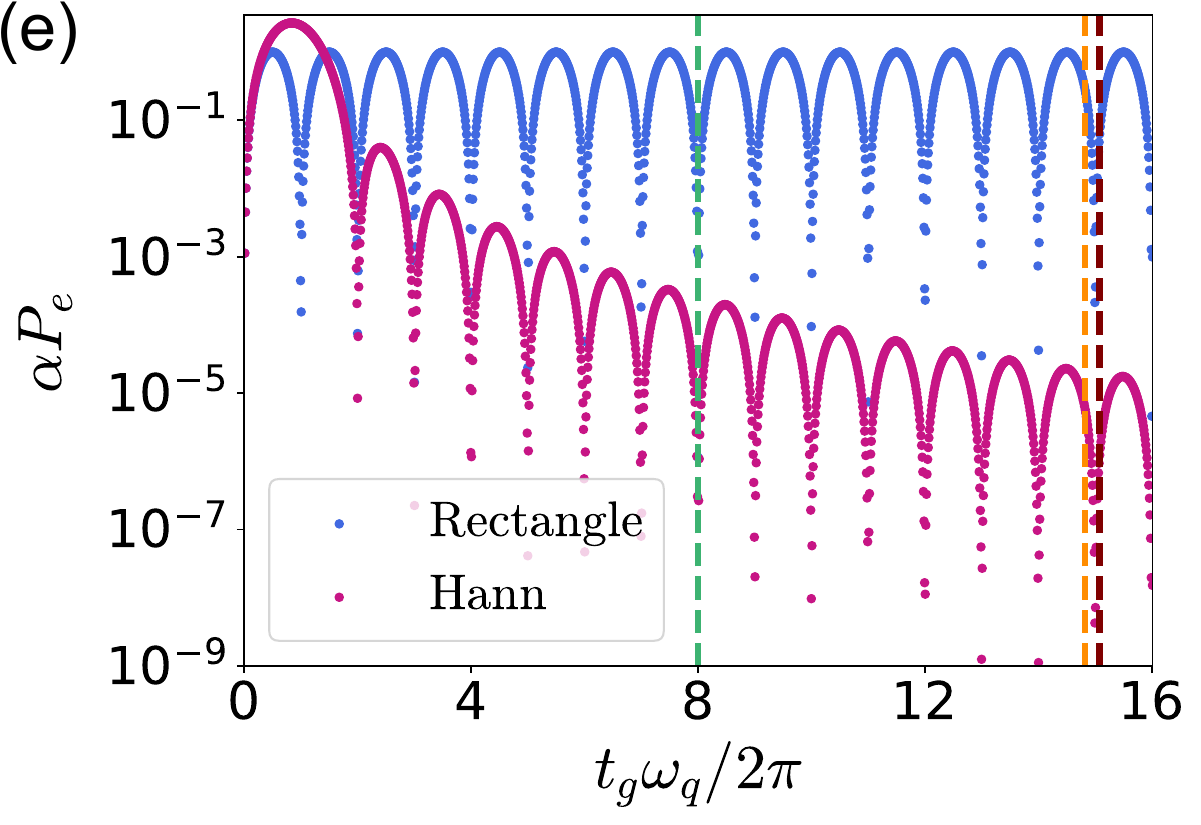}
    \label{fig:3-e}
\end{subfigure}
\caption{Short-range CZ gate in the TQD module through fast adiabatic evolution. (a) CZ gate infidelity vs time for a gate time of $t_g=100~\mathrm{ns}$ (light orange shading). Results for two values of the effective resonator-mediated coupling $J_r/h$ to the spectator qubit (qubit $D$) are shown. (b) Temporal profile of the tunnel coupling when a Hann window function is used for the CZ gate in (a) with $ \lvert J_r/h \rvert=1.68~\mathrm{kHz}$. (c) CZ gate infidelity for synchronized evolutions with $t_g \omega_q/2\pi=8$ for Hann, rectangular, and filtered rectangular pulses. (d) CZ gate infidelity for evolutions with $ t_g=100~\mathrm{nsec}$ for Hann, rectangular, and filtered rectangular pulses. (e) Normalized non-adiabatic error spectral profile for rectangular and Hann window functions. The red and orange dashed vertical lines indicate the parameter regime for the results in (a) for $\vert J_r\rvert /h=1.68$ and $10.01~\mathrm{kHz}$, respectively. The green dashed line shows the parameter regime of part (c). All gate fidelities are optimized over local operations on all qubits at each time step. Additional system parameters for (a-d) are $\omega_r/2\pi =6~\mathrm{GHz}, \omega_T^z/2\pi =5.94~\mathrm{GHz}, \omega_D^z/2\pi =5.96~\mathrm{GHz}, \omega_3^z/2\pi =5.8~\mathrm{GHz}, \lvert t_D/2\pi \rvert=\lvert t_{T 12}/2\pi \rvert=3.5~\mathrm{GHz}, t_{T 23}= t_{T 23}(t), g^x_T/2\pi=g^x_D/2\pi =200~\mathrm{MHz}, g_D^{AC}/2\pi =g_T^{AC}/2\pi =50~\mathrm{MHz}, \epsilon_{T3}/2\pi =-300~\mathrm{GHz}, U_{T1}/2\pi = U_{T2}/2\pi = U_{T3}/2\pi =2.5~\mathrm{THz}$. $J_r/h=1.68$~kHz in panels (b-d).}
\label{fig:3}
\end{figure*}

After choosing this temporal profile for the exchange coupling, in the next step, we use numerical interpolation techniques to arrive at the required inter-dot tunnel coupling $t_{T23}(t)$, based on the effective Hamiltonian values. A sample $t_{T23}(t)$ with a Hann window function is shown in Fig.~\ref{fig:3}(b). For all gate analyses in this work, we numerically solve the Schrödinger equation with the Hamiltonian in Eq.~\eqref{eq:H-Hubbard-1}. We then project the time evolution at each time step onto the low-energy subspace, effectively tracing out the resonator and higher-energy orbital degrees of freedom to arrive at the evolution of the three-qubit computational subspace. 

We quantify the performance of multi-qubit gates using the average gate fidelity $F=\frac{1}{d(d+1)}(\text{Tr}[UU^\dagger]+{ \lvert \mathrm{Tr}[{U^\dagger U_\mathrm{goal}}] \rvert}^2 )$~\cite{pedersen2007fidelity}, in which $d$ is the dimension of the Hilbert space on which the gate acts, $U$ is the actual evolution operator of the system, and $ U_\mathrm{goal}$ is the target evolution. In principle, entangling operations between two qubits in multi-qubit processors should not affect the rest of the qubits in the system, i.e., the evolution operator for the rest of the qubits should be either trivial or merely in the form of local operations which can be compensated after the two-qubit operation. To capture this crucial concept in our three-qubit system, in all gate analyses throughout this work, the reported two-qubit (CZ or CNOT) gate fidelities refer to three-qubit gate fidelities relative to a block-diagonal target gate that includes the ideal two-qubit gate in one block and a single-qubit operation on the other, i.e., $U^\mathrm{CZ}_\mathrm{goal}=\mathrm{CZ}_{T3} \otimes I_D$ for this section and $U^\mathrm{CNOT}_\mathrm{goal}=\mathrm{CNOT}_{TD} \otimes I_3$ for Section~\ref{sec-l1:LR-entanglements}. 
In trying to realize this target operation, we minimize the degree to which the two-qubit gate affects the quantum information encoded in the third qubit.

Figure~\ref{fig:3}(a) shows the infidelity ($1-F$) of a fast adiabatic CZ gate between qubits $T$ and $3$ and local operations on qubit $D$ as a function of time for a pulsed exchange coupling $J_e$ with a Hann window function profile and 100~ns gate time. Results for two different values of the coupling to the spectator qubit, which is qubit $D$ in this case, are shown. In particular, we consider $\lvert J_r\rvert /h=1.68$ and $10.01~\mathrm{kHz}$, which are calculated from the effective Hamiltonian model for two detuning values, $\epsilon_{T}/2\pi =\epsilon_{D}/2\pi =-15$ and $-10.5~\mathrm{GHz}$, respectively. The gate fidelities shown in the figure are optimized over local operations. The temporal profile of the $\lvert t_{T23}(t) \rvert$ tunnel coupling for $ \lvert J_r\rvert/h=1.68~\mathrm{kHz}$ is shown in Fig. ~\ref{fig:3}(b) for $J_0/h=10~\mathrm{MHz}$. The temporal extents of the applied control signals, $t_{T23}(t)$, are marked with a light orange background on all infidelity plots.

Here the infidelities are found to be better than $1.86\times10^{-4}$ and $9.68\times10^{-4}$ in the two cases, respectively, from which it can be concluded that by adjusting the detunings of the DQDs to within the range quoted above, the effect of the spectator qubit on the short-distance CZ gate is negligible. The CZ gate infidelities exhibit oscillatory behavior after the pulse time, $t_g$, and the infidelities given above correspond to the highest infidelities observed for all cases reported in this section. The unitarity of the evolution in the logical subspace is also better than $0.9998$ and $0.9989$, respectively. Similar CZ gate implementations with the Hann window function have been recently employed experimentally in two-qubit processors~\cite{xue2022quantum, takeda2022quantum} with very high fidelities ($99.65\%$ for~\cite{xue2022quantum}) observed.

Higher-order Fourier-basis window functions, $J_e(t)=J_0 \sum_{n}\lambda_n [1-\cos(2\pi n t/t_g)]$ may also be used for spectral engineering, where the Hann window function corresponds to the special case of a first-order window. The higher-order functions can also be optimized to minimize the non-adiabatic error for any time larger than a chosen time and have power densities that decrease with increasing order. Since these Fourier-basis windows are natural extensions of the Hann window and can potentially further reduce the error, we also analyzed second and fourth-order Fourier-basis windows with optimized Fourier coefficients $\lambda_n$~\cite{martinis2014fast} for the implementation of CZ gates for the same system parameters as in Fig.~\ref{fig:3}(a). We find an almost negligible (less than $5\times 10^{-7}$) change in the infidelities for both higher-order windows compared to the Hann window, indicating that there is little benefit to using the higher-order functions. This is due to the fact that here $t_{g}\omega_q/2\pi \approx 15.08$, and so non-adiabatic error rates are essentially overshadowed by the infidelities due to the presence of the spectator qubit as well as the coupling to the non-empty resonator states and higher orbitals. 

Though it may seem to be favorable to work in the regime with higher $t_g\omega_q/2\pi$~(see Fig.~\ref{fig:3}(e)) to reduce coherent errors due to SWAP-like evolutions, for the fixed difference in transition frequencies determined from the experiment, this condition translates to longer gate times and consequently greater decoherence. On the other hand, reducing the gate time requires larger exchange couplings, which may not be feasible in some experiments. The connection between the non-adiabatic errors and the spectral profile of the pulse enables one to design pulses that match the zeros of the power spectral density function to significantly reduce the non-adiabatic error via synchronization. We next examine this approach in the presence of the spectator qubit, while also considering the finite bandwidth of the control electronics.

The error spectra of the Hann and rectangular window functions have zeros at $t_g=2\pi m/\omega_q$ ($m \in \mathbb{Z}_+$ for rectangular and $m \in \mathbb{Z}_+\backslash \{1\}$ for Hann windows), which may be synchronized with the CZ gate time to reach a phase of $(2n+1)\pi$ in the odd-parity subspace. This results in a gate time of
\begin{equation}\label{eq:t_g-CZ-sync }
            t_g^\mathrm{sync}=\frac{\pi}{\Delta\omega}\sqrt{4 m^2-(2 n+1)^2}.
\end{equation}
Figure~\ref{fig:3}(c) shows the CZ gate infidelities for $t_g \omega_q/2\pi=8$, which is matched with a gate time of $t_g=52.98~\mathrm{ns}$ for the Hann and rectangular window functions, as well as for a low-pass-filtered rectangular function, for a spectator qubit coupling of $\lvert J_r\rvert/h=1.68~\mathrm{kHz}$. A Butterworth filter of order six is used here, with a cutoff frequency of $100~\mathrm{MHz}$ to filter the $t_{T23}(t)$ function. The green dashed line in Fig.~\ref{fig:3}(e) indicates the region associated with these pulses, with the maximum value of the exchange coupling being less than $20~\mathrm{MHz}$, consistent with the recently reported~\cite{xue2022quantum} range of $100~\mathrm{kHz}$ to $20~\mathrm{MHz}$. For the Hann, rectangular, and filtered rectangular pulses, the oscillatory infidelity functions after the pulse reach maximum infidelities of $1.84 \times 10^{-4}$, $1.92 \times 10^{-4}$, and $6.04 \times 10^{-4}$, respectively. A key observation is that for the rectangular pulses, fidelities as good as those achieved for longer CZ gates (Fig.~\ref{fig:3}(a)) may be reached, yet for much shorter pulses of time $t_g=52.98~\mathrm{ns}$. However, by limiting the bandwidth of the pulse, the gate infidelity increases by approximately a factor of three, as we can see by comparing results for the filtered and unfiltered rectangular functions.

Next, to distinguish between the effect of non-adiabatic error and bandwidth limitations on the control signals, we examine the CZ evolution for a longer gate time of $t_g=100~\mathrm{ns}$ for different pulse shapes. The results are shown in Fig.~\ref{fig:3}(d), where infidelities as low as $1.86\times10^{-4}$, $2.18\times10^{-4}$, and $2.82\times10^{-4}$ are achieved for Hann, rectangular, and filtered rectangular pulses, respectively. Comparing Hann and rectangular pulses and with the results of Fig.~\ref{fig:3}(a) and Fig.~\ref{fig:3}(c), it can be concluded that the non-adiabatic errors are no longer the limiting factor in the range of $t_g\gtrsim 100~\mathrm{ns}$ pulses. Also, we find that low-pass filtering of the pulse adds an additional infidelity penalty of $6.4\times10^{-5}$. Additionally, we conclude that with this amount of coupling to the spectator qubit, operations with Hann or perfect rectangular pulses, for both synchronized short and long gate times lead to roughly similar levels of infidelity. On the other hand, for short pulses without the synchronization condition satisfied, rectangular pulses lead to much higher infidelities compared to Hann pulses, as can also be concluded from Fig.~\ref{fig:3}(e). For example, for an asynchronized evolution of $t_g \omega_q/2\pi=8.5$ corresponding to a gate time of $t_g=56.3~\mathrm{ns}$, the rectangular pulse only reaches $1.74\times10^{-3}$ infidelity, while the Hann pulse still gives the same $1.83\times10^{-4}$ level of infidelity. Therefore, the importance of using window functions for shorter gate times, even if high bandwidth pulses are available, is undeniable. Despite all the advantages in mitigating non-adiabatic errors, Hann window functions have a big disadvantage. They require twice the inter-dot coupling strength at the maximum point compared to the rectangular pulses. Therefore, limitations on available coupling strengths may make it hard to experimentally test these shorter pulses. 


\section{\label{sec-l1:LR-entanglements}Long-range entangling gates}

A resonator-mediated CNOT gate can be implemented by applying a microwave drive to modulate the energy levels of the DQD module to create a significant interaction with qubit $T$ in the TQD. In this scheme, qubit $D$ is the control and qubit $T$ is the target of the CNOT gate, with qubit $3$ being the spectator. This protocol has been previously employed theoretically~\cite{warren2021robust} in QD systems with two spins. In this section, we review this technique and apply it in the case of a three-qubit system, enabling us to also study the effect of the spectator qubit. 

\begin{figure}[t]
\centering
\begin{subfigure}{}
    \centering \includegraphics[width=0.44\textwidth]{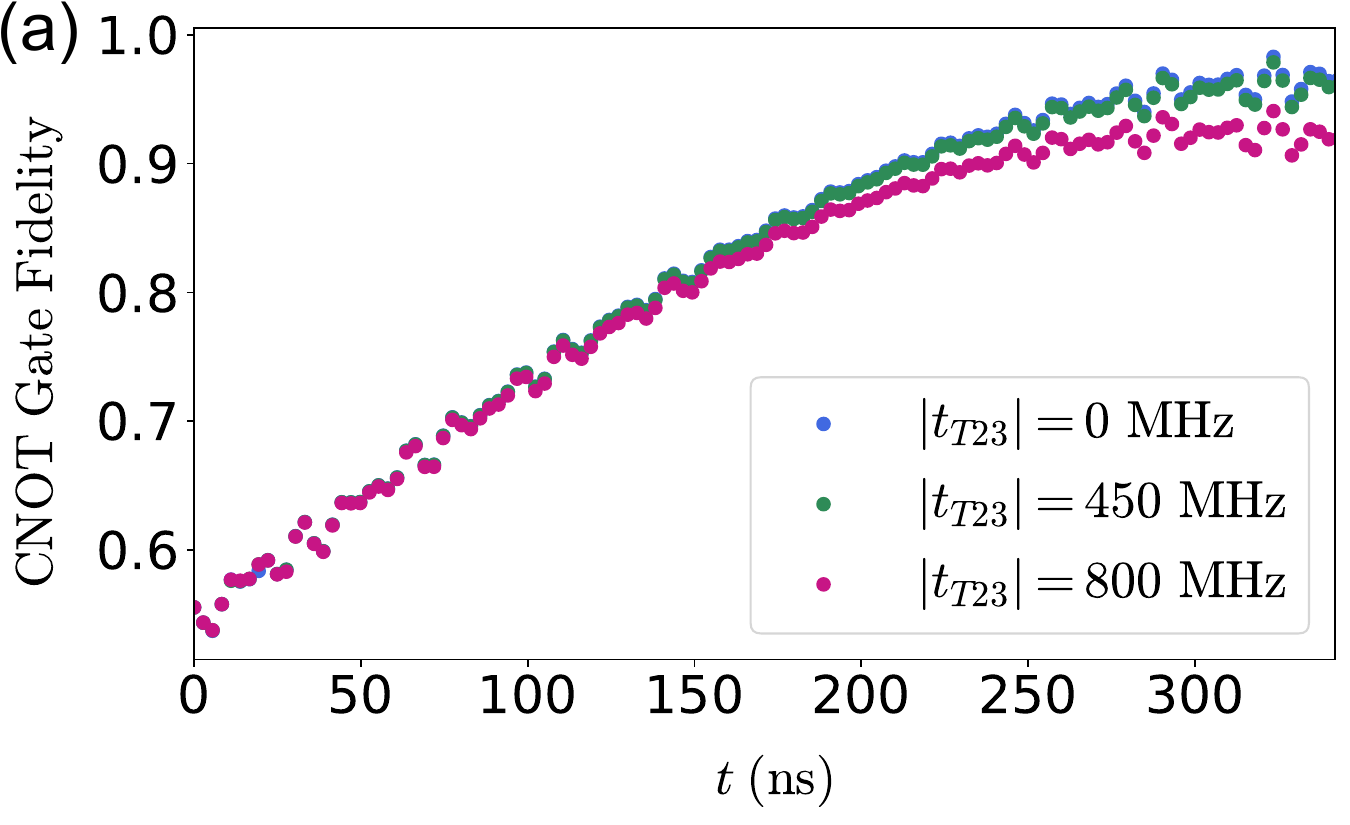}
    \label{fig:4-a}
\end{subfigure}
\\
    \vspace{0.6cm}
\begin{subfigure}{}
    \centering \includegraphics[width=0.44\textwidth]{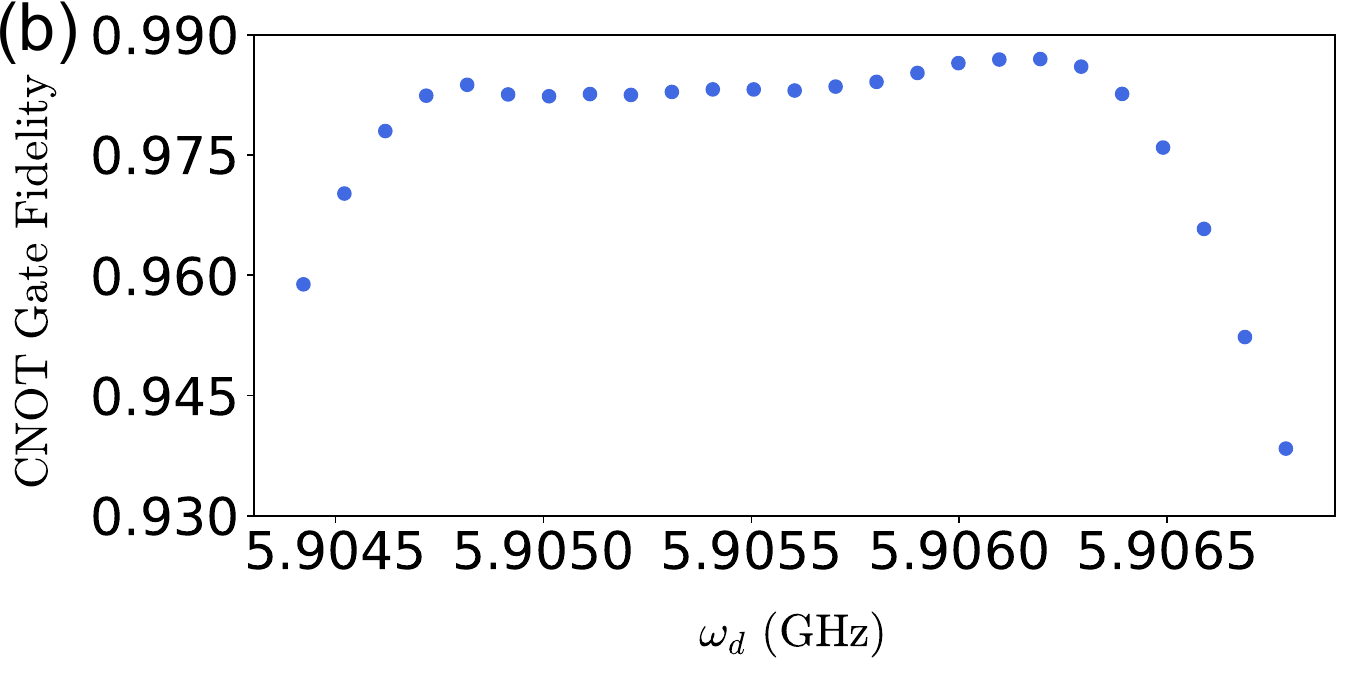}
    \label{fig:4-b}
\end{subfigure}
\caption{(a) Cross-resonance CNOT gate fidelity between qubits $D$ and $T$ vs time, optimized over local operations on all qubits at each time step. Parameters are set at $\omega_r/2\pi=6~\mathrm{GHz}, \omega_T^z/2\pi =5.94~\mathrm{GHz}, \omega_D^z/2\pi =5.96~\mathrm{GHz}, \omega_3^z/2\pi =5.8~\mathrm{GHz}, \lvert t_D/2\pi \rvert=\lvert t_{T 12} /2\pi \rvert=3.5~\mathrm{GHz}, t_{T 23}/2\pi =0, 0.45, 0.8~\mathrm{GHz}, g^x_T/2\pi =g^x_D/2\pi =200~\mathrm{MHz}, g_D^{AC}/2\pi =g_T^{AC}/2\pi =50~\mathrm{MHz}, \epsilon_{D} =\epsilon_{T}=0, \epsilon_{T3}/2\pi =-300~\mathrm{GHz}, U_{T1}/2\pi = U_{T2}/2\pi = U_{T3}/2\pi =2.5~\mathrm{THz}$. The microwave drive frequency is set to $\omega_d=\omega_T$ for cross-resonance operation. The microwave drive amplitude, $\Omega_d $, is set to $\Omega_{\mathrm{eff}}^x/2\pi =20~\mathrm{MHz}$. (b) Robustness of the gate fidelity to the cross-resonance microwave drive frequency.}
\label{fig:4}
\end{figure}

The microwave drive added to the DQD module results in the effective two-qubit low-energy Hamiltonian $\hat{H}^{\mathrm{drive}}_{\mathrm{eff}}= \cos \left(\omega_d t\right)\left(\Omega_{\mathrm{eff}}^z \sigma_D^z+\Omega_{\mathrm{eff}}^x \sigma_D^x\right)+\cos \left(2 \omega_d t\right)\left(\xi_D^z \sigma_D^z+\xi_D^x \sigma_D^x\right)+\sin \left(2 \omega_d t\right) \xi_D^y \sigma_D^y $ in the laboratory frame, with $\Omega_{\mathrm{eff}}^x =\sin\left(\beta_D\right) \Omega_D$ and $\Omega_{\mathrm{eff}}^z =\xi_D^x=\xi_D^y=\xi_D^z=0$ for zero detunings, $\epsilon_T=\epsilon_D=0$~\cite{warren2021robust}. Next, by applying a frame transformation, $U_1=\exp [-i t \omega_d\left(\sigma_D^z+\sigma_T^z\right) / 2]$, the Hamiltonian can be moved to the doubly rotating frame, and the rotating wave approximation can be made to remove the rapidly oscillating terms from the Hamiltonian, followed by another transformation using $U_2=\exp [-i \chi \sigma_D^y / 2]$, in which $\chi=\arctan \left(\Omega_{\mathrm{eff}}^x / \delta_D\right)$ and $\delta_D=\omega_D-\omega_d$, to arrive at the Hamiltonian in the diagonalized doubly rotating frame. Next, we transfer to the quadruply rotating frame using another time-dependent frame transformation, $U_3=\exp [-i t\left(\eta \sigma_D^z+\delta_T \sigma_T^z\right) / 2]$ with $\eta=\sqrt{\delta_D^2+\left(\Omega_{\mathrm{eff}}^x\right)^2}$ and $\delta_T=$ $\omega_T-\omega_d$~\cite{warren2021robust}. The final form of the Hamiltonian in this quadruply rotating frame is
\begin{equation}\label{eq:H_QF_CRCNOT}
    \begin{aligned}
            H_{\mathrm{QF}}= & -J\left(\cos ( \omega_d t ) \left\{\sin (\chi) \sigma_D^z+\cos (\chi)\left[\cos(\eta t) \sigma_D^x\right.\right.\right. \\
	& \left.\left.\left.-\sin (\eta t) \sigma_D^y\right]\right\}-\sin \left(\omega_d t\right)\left[\cos (\eta t) \sigma_D^y+\sin (\eta t) \sigma_D^x\right]\right) \\
	& \times\left[\cos \left(\omega_T t\right) \sigma_T^x-\sin \left(\omega_T t\right) \sigma_T^y\right].
     \end{aligned}
\end{equation}

We will utilize a cross-resonance interaction that results from driving the DQD energy levels in resonance with the transition frequency of  qubit $T$, $\omega_d=\omega_T$. For sufficiently small resonator-mediated interaction strength, $J_e \ll \eta$, the rotating wave approximation can be applied again to discard rapidly oscillating terms in Eq.~\eqref{eq:H_QF_CRCNOT} and arrive at a Hamiltonian of the form $H_{\mathrm{QF}} \approx-(1/2) \tilde{J_r} \sigma_D^z \sigma_T^x$, with the $ZX$ coupling coefficient $\tilde{J_r}=J_r \sin(\chi)=J_r \Omega_{\mathrm{eff}}^x / \eta$. Evolution under this Hamiltonian for time $t_g$ with $\tilde{J_r} t_g=\pi/2$ with local operations on the $T$ and $D$ qubits as $U_\mathrm{local}=\exp({i \frac{\pi}{4}}) \exp({-i \frac{\pi}{4} I_D \otimes \sigma_T^x}) \exp({-i \frac{\pi}{4} \sigma_D^z \otimes I_T})$ corresponds to a CNOT gate between the qubits.

We select the transition frequencies of the qubits to be well-detuned such that the two-qubit coherent interactions are suppressed without the microwave drive applied. Also, in tunnel-coupled QDs, the effect of charge noise can be intrinsically mitigated by running the device at the sweet spot or symmetric operation point, where the exchange coupling is first-order insensitive to the charge noise~\cite{reed2016reduced, bertrand2015quantum, martins2016noise}. Here, since the entanglement is induced only during the pulse time (with rectangular profile here) the base value of energy level detunings for the DQD and TQD are conveniently chosen to coincide with the sweet spot $\epsilon_T=\epsilon_D=0$ to suppress the charge noise at first order.

Figure~\ref{fig:4}(a) shows the cross-resonance CNOT gate fidelity between qubits $D$ and $T$ and local operations on qubit $3$ as a function of time for different values of the tunnel coupling to the spectator qubit. The figure also shows the sensitivity of the gate fidelity to the frequency of the applied microwave drive, in the absence of the spectator qubit. We now explain these results in detail.

The fidelity is calculated by solving the Schrödinger equation for the full system (Eq.~\eqref{eq:H-Hubbard-1}) with the resonator mode truncated to three photonic states, and then projecting the unitary evolution onto the three-qubit logical subspace. The maximum CNOT gate fidelity in the absence of the spectator qubit, i.e., $\lvert t_{T23}/2\pi \rvert=0.0~\mathrm{GHz}$, is $0.9829$, achieved for the gate duration of $t_g=323.6~\mathrm{ns}$. The unitarity of the evolution in the computational subspace during the gate varies between $93.34\%$ and $100\%$, which is mainly due to the leakage out of the subspace to the photon-populated resonator states during the gate and is ultimately responsible for the remaining infidelity.

In principle, by increasing the barrier height between the neighboring dots, it is possible to lower the exchange strength $J_e/h$ and nearly isolate the neighboring qubits. However, in practice, there is a limit to such isolation resulting in residual exchange interactions in the range of a few tens of $\mathrm{kHz}$ to a few hundred $\mathrm{kHz}$~\cite{xue2022quantum, mills2022two, takeda2022quantum}. We examine the effect of such residual coupling, i.e., the presence of a spectator qubit, on the fidelity of long-range entangling gate operations. 	

We consider nonzero tunnel coupling strengths of $\lvert t_{T 23}/2\pi\rvert =450~\mathrm{MHz}$ and $800~\mathrm{MHz}$, in Eq.~\eqref{eq:H-Hubbard-1}. Starting from the Hubbard model for two neighboring dots with detuning $\epsilon$ and tunnel coupling $t_c$ and considering conditions of $\lvert \epsilon \rvert < U$ and $ \vert t_c \rvert \ll U\pm \epsilon$, the low-energy hybridized singlet and triplet states are separated by $ J=\frac{4Ut_c^2}{U^2-\epsilon^2}+O \left(\frac{t_c^3}{\left(U\pm\epsilon\right)^3}\right)$, which is the Heisenberg exchange coupling strength~\cite{ scarola2004chirality, burkard2021semiconductor}. Therefore, the nonzero tunnel couplings translate to the residual coupling strength to the spectator qubit of $J_e/h=328.7~\mathrm{kHz}$ and $1.039~\mathrm{MHz}$, resulting in the maximum gate fidelity dropping to $0.9776$ and $0.9408$, respectively. 

Note that the simulated residual tunnel couplings are chosen such that clear gate fidelity degradations may be observed, while recent experimental efforts report the ability to reduce the residual exchange couplings down to a $T_2$ limit of $J_e/h \sim 10~\mathrm{kHz}$~\cite{mills2022two}, which would practically cause negligible fidelity degradations, much closer to the blue curve in Fig.~\ref{fig:4}(a). Hence, we can conclude that the presence of the third qubit would have a negligible effect on the cross-resonance CNOT gate, and increasing the barrier gate voltage between the second and third dot of the TQD module efficiently isolates the subspace the of $T$ and $D$ qubits.

As mentioned, the cross-resonance approach relies on matching the frequency of the applied drive to the transition frequency of the other qubit. In the following, we numerically examine the robustness of the gate performance to this matching condition (Fig.~\ref{fig:4}(b)). For each drive frequency, the gate duration is adapted to give the highest fidelity. It can be seen that with the commonly realized frequency resolution of signal generators, securing a cross-resonance operation should not be an issue.

The main source of qubit decoherence greatly depends on the details of the device fabrication even in isotopically enriched $^{28}\mathrm{Si}$. A key observation has been that for thin ($\sim$5nm) quantum wells such as single-layer etch-defined gate electrode devices, without magnetic field gradients, the hyperfine effect from magnetic nuclei limits the coherence time~\cite{weinstein2023universal}, while for thicker ($\sim$30-50nm) quantum wells with micromagnets fabricated on top of the device, similar to our platform, isotopic purification of the semiconductor leaves charge noise as the main source of decoherence~\cite{yoneda2018quantum, kawakami2016gate, eng2015isotopically}. Specifically, in $^{28}\mathrm{Si/SiGe}$ with extrinsic spin–electric-coupling fields applied, studying rapid spin rotations showed that the dephasing is primarily caused by charge noise and not magnetic hyperfine noise, especially since the movement of spins through local magnetic field gradients, e.g., in EDSR, makes the qubits further susceptible to electrical fluctuations~\cite{yoneda2018quantum, kawakami2016gate}. 

\begin{figure}[t]
    \centering
    \includegraphics[width = 0.95 \columnwidth]{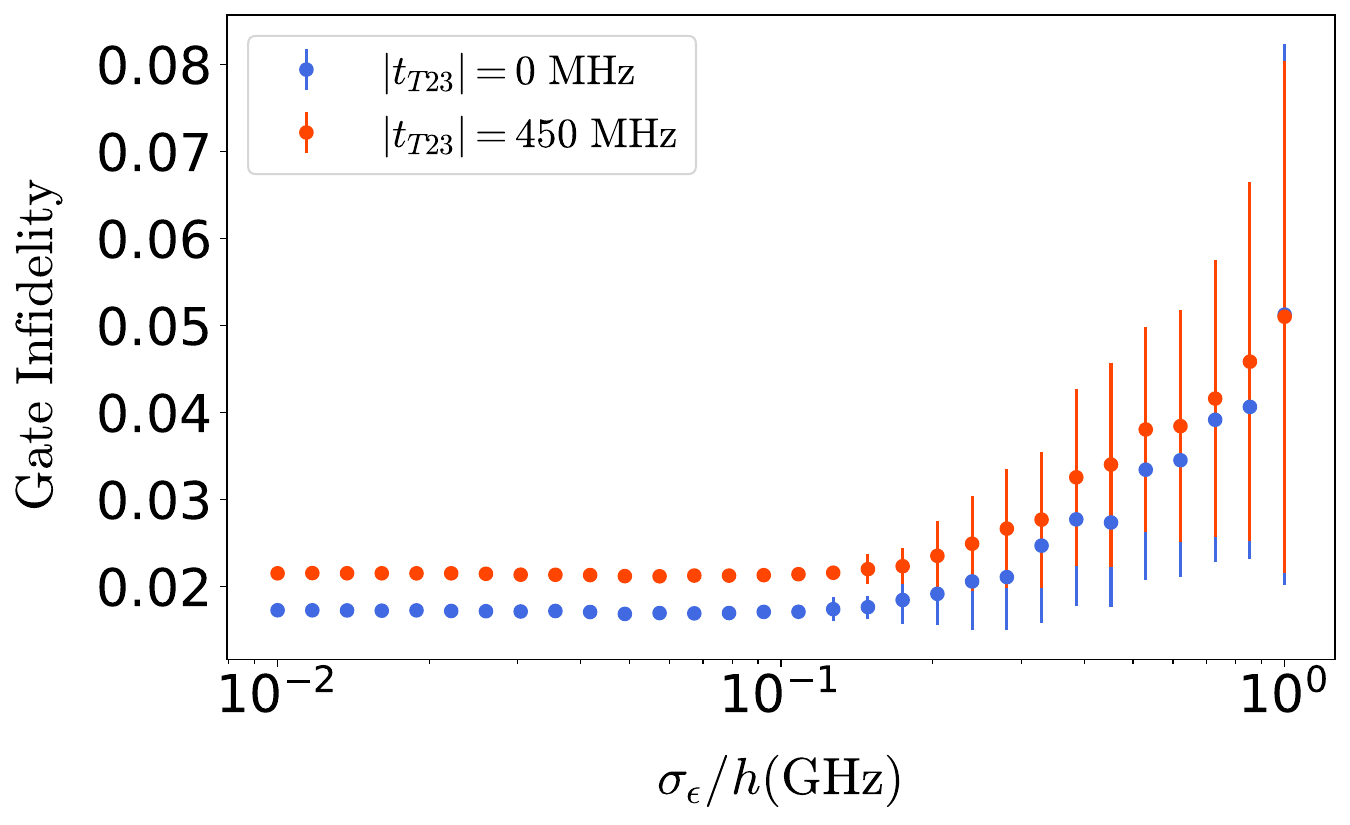}
    \caption{Cross-resonance CNOT gate infidelity between qubits $D$ and $T$ versus quasistatic charge noise amplitude, optimized over local operations on all qubits at each time step. Parameters are set at $\sigma_{t_D}=\sigma_{t_{Tij}}=\sigma_{\epsilon_i} /200=\sigma_{\epsilon_{T3}}/200$, $\omega_r/2\pi=6~\mathrm{GHz}, \omega_T^z/2\pi =5.94~\mathrm{GHz}, \omega_D^z/2\pi =5.96~\mathrm{GHz}, \omega_3^z/2\pi =5.8~\mathrm{GHz}, \lvert t_D/2\pi \rvert=\lvert t_{T 12} \rvert=3.5~\mathrm{GHz}, t_{T 23}/2\pi =0, 0.8~\mathrm{GHz}, g^x_T/2\pi =g^x_D/2\pi =200~\mathrm{MHz}, g_D^{AC}=g_T^{AC}/2\pi =50~\mathrm{MHz}, \epsilon_{D}=\epsilon_{T}=0, \epsilon_{T3}/2\pi =-300~\mathrm{GHz}, U_{T1}/2\pi = U_{T2}/2\pi = U_{T3}/2\pi =2.5~\mathrm{THz}$. The microwave drive frequency is set to $\omega_d=\omega_T$ for cross-resonance operation, and the amplitude, $\Omega_d$, is $\Omega_{\mathrm{eff}}^x/2\pi =20~\mathrm{MHz}$. A hundred sets of samples per point from the Gaussian distribution for each noise parameter were taken, with dots showing the average gate infidelity and error bars showing the standard deviation of the infidelities.}\label{fig:5}
\end{figure}

In view of this, next, we computationally study the effect of the charge noise on the cross-resonance CNOT gate fidelity through a sampling technique. We model the effect of the quasistatic charge noise by including classical fluctuations to the elements of the Hamiltonian affected by electric fields, in particular the chemical potential of dots and the tunnel couplings. Specifically, we substitute $\epsilon_i$ with $\epsilon_i+\delta \epsilon_i$, $t_D $ with $t_D +\delta t_D$, $t_{Tij} $ with $t_{Tij} +\delta t_{Tij} $, and $\epsilon_{T3}$ with $\epsilon_{T3}+\delta \epsilon_{T3}$, using Gaussian distributions for the random variables $\delta \epsilon_i, \delta t_D, \delta t_{Tij}$, and $\delta \epsilon_{T3}$ with standard deviations $\sigma_{\epsilon_i}, \sigma_{t_D}, \sigma_{t_{Tij}}$ and $\sigma_{\epsilon_{T3}}$, respectively. The inclusion of these fluctuations in the dynamics of the system results in shifts of the effective single-body and interaction parameters and potentially limits the gate fidelities. In two-qubit processors, gate sequences have been devised to suppress the sensitivity of two-qubit entangling gates to similar quasistatic charge noise effects~\cite {warren2021robust}. 

Noting that the rate of the tunnel coupling change through changing the barrier gate voltage (i.e., lever arm) is typically two orders of magnitude smaller than the rate of change for the voltage-controlled detuning levels (see \cite{borjans2021probing} and \cite{wu2014two, shi2013coherent, zajac2016scalable} for tunnel coupling and detuning control calibrations, respectively), and assuming comparable levels of voltage noise in all gates, the amplitude of detuning noise may be considered to be at least two orders of magnitude larger than the amplitude of tunnel coupling noise~\cite{reed2016reduced, martins2016noise}. Figure~\ref{fig:5} shows the infidelity of the cross-resonance CNOT gate versus the standard deviation of the detuning charge noise. For each value of the noise amplitude, a hundred sets of samples were taken from the Gaussian distributions.	 

Our analysis shows that charge noise levels smaller than $100~\mathrm{MHz}$ have negligible effect on the gate fidelity, in the presence or absence of the spectator qubit. On the other hand, for larger values of noise amplitude, the charge noise limits the fidelity of the gate and eventually dominates the infidelity due to the spectator qubit at a noise strength of $1~\mathrm{GHz}$. The level of charge noise in a given experiment highly depends on the properties of the processor and the details of the experiment. Yet, recent experiments on devices with similar geometries report charge noise values that are regularly within the left-half side of the plot~\cite{petersson2010quantum, yoneda2018quantum, mi2018landau, paquelet2023reducing, mi2018coherent}, in the region where the infidelity is unperturbed by the noise. Therefore, we do not expect charge noise to have major deleterious effects on the performance of the resonator-mediated cross-resonance CNOT gate in comparison to the infidelities induced by leakage.

\section{\label{sec-l1:Conclusion}Conclusions}

We examined multi-qubit operations in modular semiconductor QD spin qubit systems with long-range interactions mediated by superconducting resonators. We presented and analyzed a three-qubit silicon QD-based system with both delocalized and confined electrons in a TQD subsystem and a delocalized electron in a DQD subsystem, where both the TQD and DQD are capacitively coupled to a single resonator mode. We demonstrated that the dynamics of the three-qubit system can be accurately modeled via single- and two-body terms with the anticipated structure of resonator-mediated coupling between distant qubits and exchange coupling between qubits in TQD, even though one of the qubits is delocalized, together with some small residual couplings. We studied short-range fast adiabatic CZ gates, realized by pulsing on exchange couplings, and obtained high fidelities by moving to the regime where the resonator-mediated interactions are suppressed, isolating the spectator qubit. Although incoherent noise sources favor shorter gate times, we showed that for experimentally realistic parameters non-adiabatic errors may be a limiting factor in reaching such faster operations and in maintaining high fidelities unless signal processing techniques are employed to reach stable operation points. We further investigated the competing effect of spectator-associated errors and the non-adiabatic errors for this case. We also investigated the performance of resonator-mediated CNOT gates that utilize a cross-resonance scheme and showed that leakage effects are the limiting factor rather than charge noise. Our results constitute an important step toward designing large QD spin qubit processors based on resonator-coupled modules containing multiple qubits. 

\begin{acknowledgments}
This work is supported by the National Science Foundation (Grant No. 2137776) and by the U.S. Army Research Office (Grant No. W911NF-23-1-0115).
\end{acknowledgments}
\onecolumngrid
\appendix
\section{Encoded Subspace and the Effective Hamiltonian Derivation} \label{appendix:A-Eff-Ham-derivation}
In this appendix, we analytically derive an effective Hamiltonian for the dynamics of the low-energy subspace of the three-qubit QD system starting from the theoretical description of the full system and present the relevant parameters. Without loss of generality, the average magnetic field in dots $1$ and $2$ of both the TQD and DQD modules are along the $z$-axis and the magnetic field gradient is purely transversal and along the $x$-axis. Here the electron gas is extended in the $yz$-plane. Following these conditions, the total Hamiltonian in Eq.~\eqref{eq:H-Hubbard-1} can be rephrased with Pauli operators defined for the spin and orbital degrees of freedom in the two modules as follows:
\begin{equation}\label{eq:H-Pauli-form-App}
    \begin{aligned}
            \hat{\tilde{H}}_{\text{tot}} = {} & \hat{H}_{r}+\hat{\tilde{H}}_{\text{DQD}}+\hat{\tilde{H}}_{\text{TQD}}+\hat{\tilde{H}}_{\text{int}} \\
            \hat{H}_{r} = {} & \omega_{r} a^{\dagger} a \\
            \hat{\tilde{H}}_{\text{DQD}} = {} & \frac{1}{2} \epsilon_D \tilde{\tau}_D^z-t_D \tilde{\tau}_D^x+\frac{1}{2} \omega_D^z \tilde{\sigma}_D^z+g_D^x \tilde{\tau}_D^z \tilde{\sigma}_D^x\\
            \hat{\tilde{H}}_{\text{TQD}} = {} & \frac{1}{2} \epsilon_T \tilde{\tau}_T^z-\left(t_T \tilde{\tau}_T^{+}+\text {H.c. }\right)+\frac{1}{2} \omega_T^z \tilde{\sigma}_T^z+g_T^x \tilde{\tau}_T^z \tilde{\sigma}_T^x+\frac{1}{2} \vec{B}_{T 3} \cdot \vec{\sigma}_3+\frac{1}{4}\left(J_0+\left(J_{\perp} \tilde{\tau}_T^{+}+\text {H.c. }\right)+J_z \tilde{\tau}_T^z\right) \vec{\sigma}_T \cdot \vec{\sigma}_3\\
             \hat{\tilde{H}}_{\text{int}} = {} & \left(a^{\dagger}+a\right)\left(g_D^{A C} \tilde{\tau}_D^z+g_T^{A C} \tilde{\tau}_T^z\right),
     \end{aligned}
\end{equation}
in which we have defined
\begin{equation}\label{eq:H-Pauli-form-params-App}
    \begin{aligned}
            \epsilon_D = {} & \mu_{D 1}-\mu_{D 2} \\
            \omega_D^z = {} & \frac{1}{2}\left(\vec{B}_{D 1}+\vec{B}_{D 2}\right)\\
            g_D^x = {} & \frac{1}{4}\left(\vec{B}_{D 1}-\vec{B}_{D 2}\right) \\
            \epsilon_T = {} & \mu_{T1}+U_{T31}-\mu_{T2}-U_{T23} \\
	      & +\frac{1}{2} t_{T23}^2\left(\frac{1} 
            {U_{T 2}-U_{T 23}+\mu_{T 2}-\mu_{T 3}}+\frac{1}{U_{T 3}-U_{T 23}-\mu_{T 2}+\mu_{T 3}}+\frac{2}{U_{T 31}-U_{T 12}-\mu_{T 2}+\mu_{T 3}}\right) \\
	      & -\frac{1}{2}\left|t_{T    
            31}\right|^2\left(\frac{1}{U_{T 1}-U_{T 31}+\mu_{T 1}-\mu_{T 3}}+\frac{1}{U_{T 3}-U_{T 31}-\mu_{T 1}+\mu_{T 3}}+\frac{2}{U_{T 23}-U_{T 12}-\mu_{T 1}+\mu_{T 3}}\right) \\
            t_T = {} & t_{T 12}+\frac{1}{4} t_{T 23} t_{T 31}\left(\frac{1}{U_{T 3}-U_{T 23}-\mu_{T 2}+\mu_{T 3}}+\frac{1}{U_{T 3}-U_{T 31}-\mu_{T 1}+\mu_{T 3}}\right. \\
	      & \left.+\frac{1}{U_{T 23}-U_{T 12}- 
            \mu_{T 1}+\mu_{T 3}}+\frac{1}{U_{T 31}-U_{T 12}-\mu_{T 2}+\mu_{T 3}}\right) \\
            \omega_T^z = {} & \frac{1}{2}\left(\vec{B}_{T 1}+\vec{B}_{T 2}\right) \\
            g_T^x = {} & \frac{1}{4}\left(\vec{B}_{T 1}-\vec{B}_{T 2}\right) \\
            J_0 = {} & \left|t_{T 31}\right|^2\left(\frac{1}{U_{T 1}-U_{T 31}+\mu_{T 1}-\mu_{T 3}}+\frac{1}{U_{T 3}-U_{T 31}-\mu_{T 1}+\mu_{T 3}}\right) \\
	      & +t_{T 23}^2\left(\frac{1}{U_{T 2}- 
            U_{T 23}+\mu_{T 2}-\mu_{T 3}}+\frac{1}{U_{T 3}-U_{T 23}-\mu_{T 2}+\mu_{T 3}}\right) \\
            J_{\perp} = {} & t_{T 23} t_{T 31}\left(\frac{1}{U_{T 3}-U_{T 23}-\mu_{T 2}+\mu_{T 3}}+\frac{1}{U_{T 3}-U_{T 31}-\mu_{T 1}+\mu_{T 3}}\right. \\
	      & \left.+\frac{1}{U_{T 12}-U_{T 
            23}+\mu_{T 1}-\mu_{T 3}}+\frac{1}{U_{T 12}-U_{T 31}+\mu_{T 2}-\mu_{T 3}}\right) \\
            J_z = {} & \left|t_{T 31}\right|^2\left(\frac{1}{U_{T 1}-U_{T 31}+\mu_{T 1}-\mu_{T 3}}+\frac{1}{U_{T 3}-U_{T 31}-\mu_{T 1}+\mu_{T 3}}\right) \\
	      & -t_{T 23}^2\left(\frac{1}{U_{T 2}- 
            U_{T 23}+\mu_{T 2}-\mu_{T 3}}+\frac{1}{U_{T 3}-U_{T 23}-\mu_{T 2}+\mu_{T 3}}\right),            
     \end{aligned}
\end{equation}
with the spin and orbital single-qubit \textit{j}th Pauli operators for the delocalized electrons in module $i$ labeled as $\tilde{\sigma}_i^j$ and $\tilde{\tau}_i^j$, respectively. $\tilde{\sigma}_3^j$ are the spin Pauli operators of the electron localized in the third dot of the TQD module. Capacitive resonator coupling and inter-dot coupling strengths are small enough compared to the energy differences of the Hamiltonian that their effect can be considered perturbatively, with the unperturbed Hamiltonian given by
\begin{equation}\label{eq:H0-unperturbed-App}
             \tilde{H}_0=\sum_{i=D, T}\left(\frac{1}{2} \epsilon_i \tilde{\tau}_i^z-\left(t_i \tilde{\tau}_i^{+}+\text {H.c. }\right)+\frac{1}{2} \omega_i^z \tilde{\sigma}_i^z+g_i^x \tilde{\tau}_i^z \tilde{\sigma}_i^x\right)+\frac{1}{2} \vec{B}_{T 3} \cdot \vec{\sigma}_3.
\end{equation}
We start by diagonalizing the unperturbed DQD Hamiltonians, which is conducted in three steps, following a similar approach as some previous theoretical works~\cite{benito2019optimized, warren2021robust}. In the first step, we diagonalize the single-body orbital terms with the unitary transformation $U_1$ defined as 
\begin{equation}\label{eq:U1-App}
             U_1=\prod_i \exp \left(i \frac{\phi_i}{2} \tilde{\tau}_T^z\right) \exp \left(-i \frac{\theta_i}{2} \tilde{\tau}_i^y\right),
\end{equation}
with parameters,
\begin{equation}\label{eq:U1-params-App}
    \begin{aligned}
            \phi_{i=D, T} = {} & \arg t_i \\
            \theta_i = {} & \arctan \left(\frac{-2\left|t_i\right|}{\epsilon_T}\right) \\
            \omega_i^a = {} & \sqrt{\epsilon_i^2+4\left|t_i\right|^2}.
     \end{aligned}
\end{equation}
Applying the $U_1$ transformation we find
\begin{equation}\label{eq:U1-applied-App}
             U_1^{\dagger} \tilde{H}_0 U_1=\sum_{i=D, T}\left(\frac{1}{2} \omega_i^a \tilde{\tau}_i^z+\frac{1}{2} \omega_i^z \tilde{\sigma}_i^z+g_i^x\left(\cos \left(\theta_i\right) \tilde{\tau}_i^z-\sin \left(\theta_i\right) \tilde{\tau}_i^x\right) \tilde{\sigma}_i^x\right)+\frac{1}{2} \vec{B}_{T 3} \cdot \vec{\tilde{\sigma}}_3.
\end{equation}
In the second step we diagonalize the $\tilde{\tau}_i^z \tilde{\sigma}_i^x$ and remaining single-body magnetic terms with a second unitary transformation $U_2$ of the form
\begin{equation}\label{eq:U2-App}
             U_2=\exp \left(-i \frac{\phi_3}{2} \tilde{\sigma}_3^z\right) \exp \left(-i \frac{\alpha_3}{2} \tilde{\sigma}_3^y\right) \prod_{i=D, T} \exp \left(-i \frac{\alpha}{2} \tilde{\tau}_i^z \tilde{\sigma}_3^y\right),
\end{equation}
with parameters,
\begin{equation}\label{eq:U2-params-App}
    \begin{aligned}
            \phi_3 = {} & \arctan \left(\frac{B_{T 3}^y}{B_{T 3}^x}\right) \\
            \alpha_{i=D, T} = {} & \arctan \left(\frac{2 g_i^x \cos \left(\theta_i\right)}{\omega_i^z}\right) \quad \alpha_3=\arccos \left(\frac{B_{T 3}^z}{\left|\vec{B}_{T 3}\right|}\right) \\
            \omega_{i=D,T}^{z^{\prime}} = {} & \sqrt{\left(\omega_i^z\right)^2+\left(2 g_i^x \cos \left(\theta_i\right)\right)^2} \quad \omega_3^\sigma=\left|\vec{B}_{T 3}\right|,
     \end{aligned}
\end{equation}
which transforms the Hamiltonian as follows,
\begin{equation}\label{eq:U1U2-applied-App}
             U_2^{\dagger} U_1^{\dagger} \tilde{H}_0 U_1 U_2=\sum_{i=D, T}\left(\frac{1}{2} \omega_i^a \tilde{\tau}_i^z+\frac{1}{2} \omega_i^{z^{\prime}} \tilde{\sigma}_i^z-g_i^x \sin \left(\theta_i\right) \tilde{\tau}_i^x \tilde{\sigma}_i^x\right)+\frac{1}{2} \omega_3^\sigma \tilde{\sigma}_3^z.
\end{equation}
The third step is transformation $U_3$ in order to eliminate the remaining two-body terms, i.e., $ \tilde{\tau}_i^x \tilde{\sigma}_i^x$, and fully diagonalize $\tilde{H}_0$.
\begin{equation}\label{eq:U3-App}
             U_3=\prod_{i=D, T} \exp \left(-i \frac{1}{2}\left(\frac{\beta_i^{+}+\beta_i^{-}}{2} \tilde{\tau}_i^y \tilde{\sigma}_i^x+\frac{\beta_i^{+}-\beta_i^{-}}{2} \tilde{\tau}_i^x \tilde{\sigma}_i^y\right)\right),
\end{equation}
with parameters,
\begin{equation}\label{eq:U3-params-App}
    \begin{aligned}
            \omega_i^\tau \pm \omega_i^\sigma = {} & \sqrt{\left(\omega_i^a \pm \omega_i^{z^{\prime}}\right)^2+\left(2 g_i^x \sin \left(\theta_i\right)\right)^2} \\
            \beta_i^{ \pm} = {} & \arctan \left(\frac{-2 g_i^x \sin \left(\theta_i\right)}{\omega_i^a \pm \omega_i^{z^{\prime}}}\right) \\
            \beta_i = {} & \frac{1}{2}\left(\beta_i^{+}+\beta_i^{-}\right).
     \end{aligned}
\end{equation}
After applying all three unitary operations, the Hamiltonian is fully diagonalized, which can be equivalently summarized via transformed Pauli operators for spins and orbitals as

\begin{equation}\label{eq:H0-transformed-App}
           \tilde{H}_0 =\omega_r a^{\dagger} a+\sum_{i=D, T} \frac{1}{2} \omega_i^\tau \tau_i^z+\sum_{i=D, T, 3} \frac{1}{2} \omega_i^\sigma \sigma_i^z,
\end{equation}
with transformed Pauli operators
\begin{equation}\label{eq:Transformed Paulis-App}
    \begin{aligned}
            \tau_i = {} & U_1 U_2 U_3 \tilde{\tau}_i U_3^{\dagger} U_2^{\dagger} U_1^{\dagger} \\
            \sigma_i = {} & U_1 U_2 U_3 \tilde{\sigma}_i U_3^{\dagger} U_2^{\dagger} U_1^{\dagger}.
     \end{aligned}
\end{equation}
Now, the total transformed system Hamiltonian is
\begin{equation}\label{eq:Total Transformed Hamiltonian-App}
    \begin{aligned}
            \tilde{H} = {} & \tilde{H}_0+\tilde{H}_I \\
            \tilde{H}_I = {} & \sum_{i=D, T} g_i^{A C}\left(a^{\dagger}+a\right) d_i+H_{\mathrm{exchg}},
     \end{aligned}
\end{equation}
where $d_i$ are transformed dipole operators, similar to the previously studied resonator-mediated two-electron systems~\cite{warren2021robust}, and $H_{\mathrm{exchg}}$ is the transformed exchange Hamiltonian:
\begin{equation}\label{eq:d_i-App}
    \begin{aligned}
            d_i & = {} \tilde{\tau}_i^z = {} \frac{1}{2}\left\{\cos \left(\theta_i\right)\left[\cos \left(\beta_i^{+}\right)+\cos \left(\beta_i^{-}\right)\right]-\sin \left(\theta_i\right) \sin \left(\alpha_i\right)\left[\sin \left(\beta_i^{+}\right)-\sin \left(\beta_i^{-}\right)\right]\right\} \tau_i^z\\
            & +\frac{1}{2}\left\{\cos \left(\theta_i\right)\left[\cos \left(\beta_i^{+}\right)-\cos \left(\beta_i^{-}\right)\right]\right. \left.-\sin \left(\theta_i\right) \sin \left(\alpha_i\right)\left[\sin \left(\beta_i^{+}\right)+\sin \left(\beta_i^{-}\right)\right]\right\} \sigma_i^z\\
            & -\frac{1}{2}\left\{\cos \left(\theta_i\right)\left[\sin \left(\beta_i^{+}\right)+\sin \left(\beta_i^{-}\right)\right]+\sin \left(\theta_i\right) \sin \left(\alpha_i\right)\left[\cos \left(\beta_i^{+}\right)-\cos \left(\beta_i^{-}\right)\right]\right\}\sigma_i^x \tau_i^x \\
	      & +\frac{1}{2}\left\{\cos    
            \left(\theta_i\right)\left[\sin \left(\beta_i^{+}\right)-\sin \left(\beta_i^{-}\right)\right]+\sin \left(\theta_i\right) \sin \left(\alpha_i\right)\left[\cos \left(\beta_i^{+}\right)+\cos \left(\beta_i^{-}\right)\right]\right\} \sigma_i^y \tau_i^y-\sin \left(\theta_i\right) \cos \left(\alpha_i\right) \cos \left(\beta_i\right) \tau_i^x \\
            & -\sin \left(\theta_i\right) \cos \left(\alpha_i\right) \sin \left(\beta_i\right) \sigma_i^x \tau_i^z .
     \end{aligned}
\end{equation}

Next, we apply a Schrieffer-Wolff transformation~\cite{schrieffer1966relation}, to isolate the dynamics governing the low-energy subsystem to leading order, treating $\tilde{H}_0$ as the unperturbed Hamiltonian and $\tilde{H}_I$ as the perturbation, under the assumption $g_i^{AC} \ll \lvert \omega_i^\tau-\omega_r \rvert,~\lvert \omega_r-\omega_i^\sigma \rvert,~\lvert \omega_r-\left(\omega_i^\tau-\omega_i^\sigma \right) \rvert$. This analytical step in combination with numerical comparison allows us to eliminate the couplings of QD modules and the resonator to leading order as well as the coupling between the ground and excited orbital states, defining the low-energy subspace as $\braket{a^\dagger a}=0$ and $\braket{\tau_i^z}=-1$. Eventually, this enables us to capture the dynamics of the low-energy encoded subspace with a good approximation via the effective Hamiltonian $H_\mathrm{eff}$,
\begin{equation}\label{eq:Heff-App}
             \hat{H}_\mathrm{eff}=\sum_{i=D, T, 3} \frac{1}{2} \omega_i \sigma_i^z-J_r \sigma_D^x \sigma_T^x+e^{i \alpha_3 \sigma_3^y} e^{i \phi_3 \sigma_3^z}\left(\frac{J_e}{4} \vec{\sigma}_T \cdot \vec{\sigma}_3+J_{Z Z} \sigma_T^z \sigma_3^z\right) e^{-i \phi_3 \sigma_3^z} e^{-i \alpha_3 \sigma_3^y}.
\end{equation}
In this effective Hamiltonian, the parameters $\phi_3$ and $\alpha_3$ are defined in Eq.~\eqref{eq:U2-params-App} and the rest of the parameters are as follows,
\begin{equation}\label{eq:Eff Hamiltonian-All Params-App}
    \begin{aligned}
            \omega_{i=D, T} = {} & \omega_i^\sigma-2 \omega_i^\sigma \frac{\left(g_i^{A C}\right)^2}{\omega_r^2-\left(\omega_i^\sigma\right)^2}\left(\sin \left(\theta_i\right) \cos \left(\alpha_i\right) \sin \left(\beta_i\right)\right)^2 \\
	      & +\left(\frac{\left(g_i^{A 
            C}\right)^2} 
        {\omega_r+\omega_i^\tau+\omega_i^\sigma}-\frac{\omega_i^\tau+\omega_i^\sigma}{2} \frac{\tilde{\Omega}_i^2}{\left(\omega_i^d\right)^2-\left(\omega_i^\tau+\omega_i^\sigma\right)^2}\right)\left(\cos \left(\theta_i\right) \sin \left(\beta_i^{+}\right)+\sin \left(\theta_i\right) \sin \left(\alpha_i\right) \cos \left(\beta_i^{+}\right)\right)^2 \\
	& +\left(\frac{\omega_i^\tau-\omega_i^\sigma}{2} \frac{\tilde{\Omega}_i^2}{\left(\omega_i^d\right)^2-\left(\omega_i^\tau-\omega_i^\sigma\right)^2}-\frac{\left(g_i^{A C}\right)^2}{\omega_r+\omega_i^\tau-\omega_i^\sigma}\right)\left(\cos \left(\theta_i\right) \sin \left(\beta_i^{-}\right)-\sin \left(\theta_i\right) \sin \left(\alpha_i\right) \cos \left(\beta_i^{-}\right)\right)^2 \\
	& +\frac{g_i^{A C}}{\omega_r}\left(\cos \left(\theta_i\right)\left(\cos \left(\beta_i^{+}\right)-\cos \left(\beta_i^{-}\right)\right)-\sin \left(\theta_i\right) \sin \left(\alpha_i\right)\left(\sin \left(\beta_i^{+}\right)+\sin \left(\beta_i^{-}\right)\right)\right) \\
	& \times \sum_{j=D, T} g_j^{A C}\left(\cos \left(\theta_j\right)\left(\cos \left(\beta_j^{+}\right)+\cos \left(\beta_j^{-}\right)\right)-\sin \left(\theta_j\right) \sin \left(\alpha_j\right)\left(\sin \left(\beta_j^{+}\right)-\sin \left(\beta_j^{-}\right)\right)\right) \\
            J_r = {} & \omega_r g_D^{A C} g_T^{A C} \sin \left(\theta_D\right) \cos \left(\alpha_D\right) \sin \left(\beta_D\right) \sin \left(\theta_T\right) \cos \left(\alpha_T\right) \sin \left(\beta_T\right)\left(\frac{1}{\omega_r^2-\left(\omega_D^\sigma\right)^2}+\frac{1}{\omega_r^2-\left(\omega_T^\sigma\right)^2}\right)\\
            \omega_3 = {} & \omega_3^\sigma-\frac{1}{4} J_0 \cos \left(\alpha_T\right)\left(\cos \left(\beta_T^{+}\right)-\cos \left(\beta_T^{-}\right)\right) \\
            J_e = {} & \cos \left(\frac{\beta_T^{-}-\beta_T^{+}}{2}\right) \cos \left(\alpha_T\right) J_0-\cos \left(\beta_T\right) \cos \left(\alpha_T\right)\left(J_z \cos \left(\theta_T\right)+\operatorname{Re}\left(J_{\perp} e^{-i \phi_T}\right) \sin \left(\theta_T\right)\right) \\
            J_{Z Z} = {} & -\frac{1}{2} \sin ^2\left(\frac{\beta_T}{2}\right) \cos \left(\alpha_T\right)\left(\cos \left(\frac{\beta_T^{-}-\beta_T^{+}}{2}\right) J_0+J_z \cos \left(\theta_T\right)+\operatorname{Re}\left(J_{\perp} e^{-i \phi_T}\right) \sin \left(\theta_T\right)\right).
     \end{aligned}
\end{equation}
For the considered arrangement of the magnetic field in the TQD module (Fig.~\ref{fig:1}), $\phi_3=0$ and $\alpha_3=0$, corresponding to the effective Hamiltonian expressed in the main text (Eq.~\eqref{eq:H-Eff-Maintext}). We tested the validity of this effective Hamiltonian numerically with the results reported in Section~\ref{sec-l2:Testing the effective model} of the main text.

\twocolumngrid


\providecommand{\noopsort}[1]{}\providecommand{\singleletter}[1]{#1}%

\end{document}